# The Impact of Credit Risk and Implied Volatility on Stock Returns

Florian Steiger[1]

*Working Paper, May 2010*

Abstract:

This paper examines the possibility of using derivative-implied risk premia to explain stock returns. The rapid development of derivative markets has led to the possibility of trading various kinds of risks, such as credit and interest rate risk, separately from each other. This paper uses credit default swaps and equity options to determine risk premia which are then used to form portfolios that are regressed against the returns of stock portfolios. It turns out that both, credit risk and implied volatility, have high explanatory power in regard to stock returns. Especially the returns of distressed stocks are highly dependent on credit risk fluctuations. This finding leads to practical implications, such as cross-hedging opportunities between equity and credit instruments and potentially allows forecasting stock returns based on movements in the credit market.

[1] Author: Florian Steiger, e-mail: florian.steiger@post.harvard.edu

# Contents

## Table of Figures

# 1 Introduction

## 1.1 Problem Definition and Objective

This paper tries to examine the possibility of using derivative-implied risk premia to explain stock returns. The intuition behind this approach is that the rapid development of derivative markets within the last ten years has led to the possibility of trading various kinds of risks, such as credit and interest rate risk, separately from each other. Also the abundance of evidence that derivative markets react faster to news than cash markets is advantageous to building an asset pricing model based on financial derivatives instead of cash market products.

This approach is different to that of most classic asset pricing models such as the CAPM and multifactor models like the Fama-French or Carhart models. These models use cash market instruments to calculate risk premia, which are then used in a multivariate regression to determine an asset's sensitivity, or factor loadings, to these risk premia.

However, parts of the methodology of the classical Fama-French three factor model are still used in the empirical part of this paper. The main difference in this paper is that the determining factors are based on financial derivatives instead of cash market instruments. To be specific, this paper uses the implied volatility based on equity options and the credit risk based on credit default swap spreads as determining risk factors in the pricing model. Ultimately, it is the objective of this paper to test if these derivative-implied risk factors and their corresponding risk premia are sufficient to explain stock returns. Hence, the statistical significance and the measures for the overall explanatory power, specifically the $R^2$, of this analysis will in the end be the indications to determine if an approach that uses derivative-implied data is successful to adequately explain stock returns.

## 1.2 Course of the Investigation

In the beginning, this paper reviews the classic asset pricing theories starting with the capital asset pricing model followed by multifactor models. Specific emphasize is being placed on the Fama-French and the Carhart models, since they use approaches in determining the risk premia and factor loadings that are similar to those used in this paper. Afterwards, the theoretic advantages of using derivative markets instead of cash markets for multivariate asset pricing models are discussed.

In the next section, implied volatility and credit risk are introduced as risk factors. Especially the determination of the implied volatility based on the Black-Scholes option pricing formula is discussed together with its underlying assumptions and constraints. Afterwards, the literature findings about possible relationships between implied volatility, credit default swap spreads, and stock returns is reviewed.

In the following empirical section, the exact model specifications and econometric approaches of this paper are introduced together with the selection criteria for the relevant stocks and derivatives which are included in this study. Afterwards the formation of the stock portfolios, against which the risk factors are later regressed, is described. Then, in the main part of this paper, the risk premia for implied volatility and the credit risk are determined. Following that, the factor loadings of the analyzed portfolios on these risk premia as well as their statistical properties are derived through a multivariate regression analysis.

The results of this regression analysis are discussed in the next section. Specific emphasize is put on the factor loadings, their statistical significance, and the overall explanatory power of this analysis. Afterwards, a short summary of the findings concludes this paper together with some ideas for further research.

# 2 Asset Pricing Models and Derivative Markets

## 2.1 The CAPM

One of the most well known and widely used models to analyze stock returns is the *capital asset pricing model* (CAPM) that was independently developed by William Sharpe (1964) and John Lintner (1965). The CAPM is based on the portfolio optimization theory by Henry Markowitz (1952), who describes portfolio selection as a question of achieving high returns while minimizing overall portfolio risk or variance. He argues that rational investors will always try to maximize their return whilst minimizing their portfolio variance. The Sharpe-Ratio, which is one of the most often used ratios in modern finance, is exactly the mathematical statement of this relationship as it is the ratio of excess return of a portfolio to its risk, as described by its standard deviation (Sharpe, 1966):

$$Sharpe-Ratio = \frac{R_i - R_f}{\sigma_i}$$

The progress of modern portfolio theory has led to the development of more sophisticated models to describe this classic risk-return relationship. The Sortino ratio for example uses one sided moments to include only the downside volatility in the denominator. Instead of calculating the excess return against the risk-free rate, the Sortino ratio uses an arbitrary threshold in the numerator (Sortino & Van der Meer, 1991). This threshold could be the minimum return guaranteed to outside investors or the maximum loss a portfolio could absorb before margin calls lead to its liquidation.

$$Sortino - Ratio = \frac{R_i - T}{\theta_i(t)}$$

where

$$\theta_i(t) = \left( \int_{\infty}^{t} \left(t - \widetilde{r_p}\right)^2 pdf(\widetilde{r_p}) d\widetilde{r_p} \right)^{0.5}$$

Even though formulas have become more complex since the first approaches were made by Markowitz, the idea of return maximization and risk minimization is still one of the underlying concepts of modern finance. Hence, the CAPM, which is based on these very fundamentals, is still of practical and theoretical use. In order to understand more advanced asset pricing models, the CAPM is quickly introduced in this section.

The CAPM is an equilibrium based model which assumes that all investors agree on the distribution of asset returns and that all investors hold efficient frontier portfolios. A portfolio on the efficient frontier is the portfolio with the highest possible return for a given standard deviation. The CAPM furthermore assumes that there is a risk-free asset paying the interest rate $R_f$ which can also be used to borrow money at the same rate. Asset demand and asset supply are in equilibrium in the CAPM.

These assumptions lead to several implications: The most important implication is that the portfolio with the highest Sharpe-Ratio, also known as the tangent portfolio, is the market portfolio.

**Picture 1: CAPM - Efficient frontier**

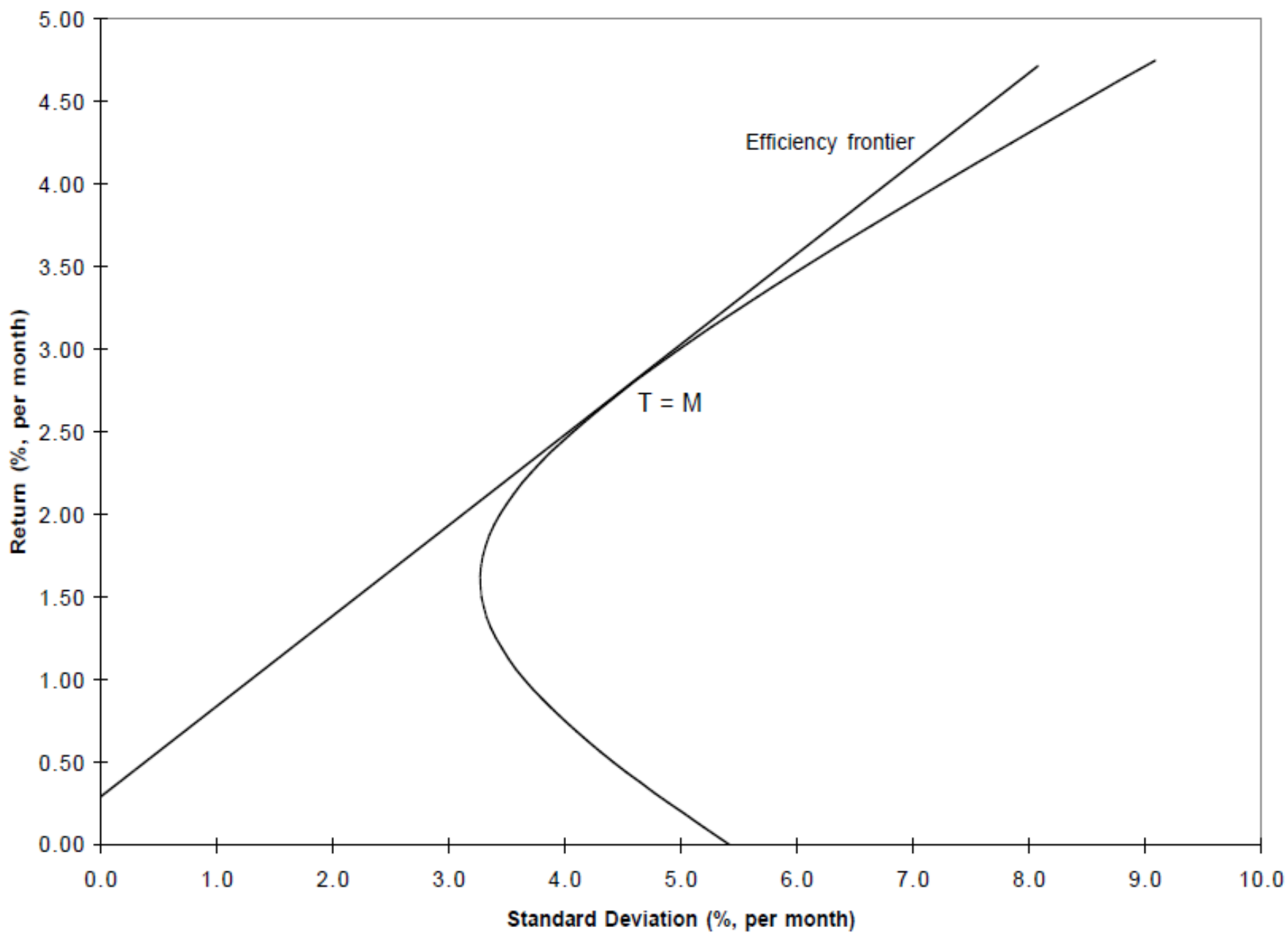

The second implication, also known as Tobin separation theorem, is that all investors hold the market portfolio. The only difference between investors with different utility functions is the overall proportion of their wealth which they invest into the market portfolio and into the risk-free asset.

The CAPM further states that the return of a risky asset only depends on this asset's sensitivity to market risk. This sensitivity is commonly known as a stock's beta. This beta can be derived as follows. In equilibrium, the marginal return-to-risk of a risky asset ($RRR_i$) must be identical to the return-to-risk of the tangent or market portfolio ($RRR_M$).

Where

$$RRR_i = \frac{\overline{r_i} - r_f}{\sigma_{iM} / \sigma_M}$$

$$RRR_M = \frac{\overline{r_M} - r_f}{\sigma_M}$$

Setting these two equal and solving for the excess return yields:

$$\overline{r_i} - r_f = \frac{\sigma_{iM}}{\sigma_M{}^2}\left(\overline{r_M} - r_f\right)$$

Substituting $\beta_i = \frac{\sigma_{iM}}{{\sigma_M}^2}$ yields:

$$\overline{r_i} - r_f = \beta_i\left(\overline{r_M} - r_f\right)$$

This formula states that a security's excess return is fully dependent on its systematic risk and that idiosyncratic risk is not being rewarded. As discussed, the beta of a security is the sensitivity to systematic risk. Securities with higher betas have higher expected returns and vice-versa. This relationship between betas and expected return is known as the *security market line* (SML).

**Picture 2: CAPM - The SML**

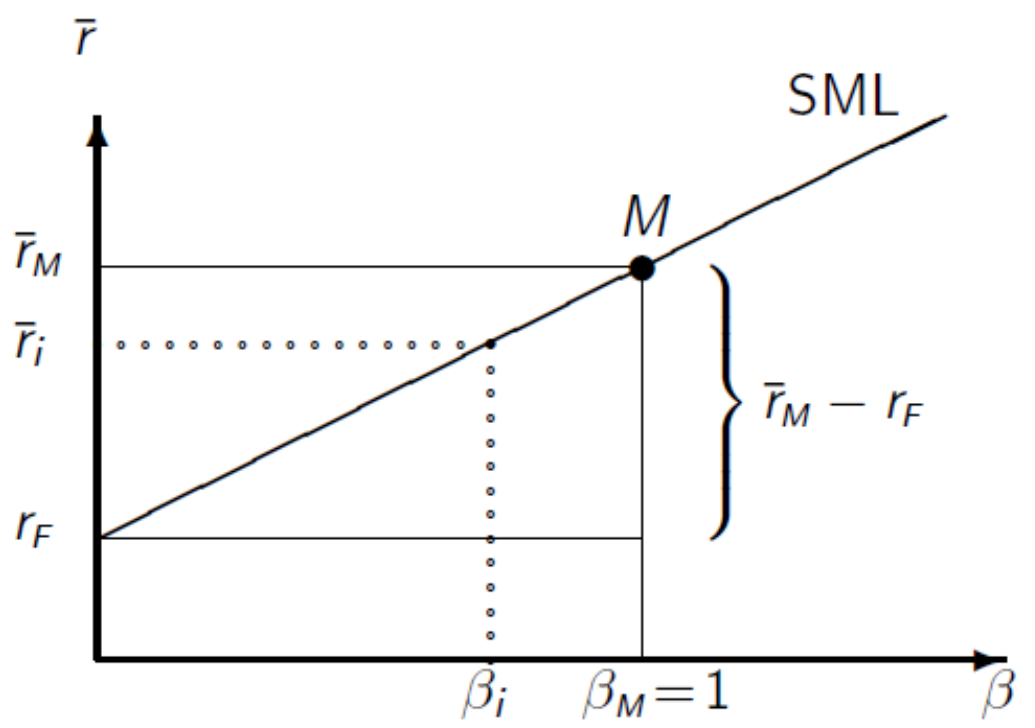

This line states that higher returns are associated with higher market betas, hence higher systematic or market risk. This relationship is linear in its nature and assets that are not exposed to any systematic or market risk have an expected return that is equal to the risk-free interest rate.

So if the CAPM holds, a security's return would be fully determined by its beta. This testable implication does not hold very well in practice. Black, Jensen, and Scholes (1972) have shown that the expected return of an asset is not strictly proportional to its $\beta$ and hence, that the CAPM does not hold perfectly. The CAPM can be empirically tested using the following equation (Black, Jensen, & Scholes, 1972, p. 4), where the alpha is the stock's abnormal return not explained by the CAPM equation:

$$\alpha = E\left(\widetilde{R_j}\right) - E(\widetilde{R_m})\beta_j$$

If the CAPM accurately described the stock's returns, alpha in this equation should not be statistically significant different from zero. In their study, Black, Jenson, and Scholes show that non-zero alphas, which are statistically significant, occur frequently throughout their sample (Black, Jensen, & Scholes, 1972, p. 16). Thus the CAPM does not perfectly hold in all cases and can be rejected as an all-explaining model of stock returns. This fact is also underlined by Merton's (1973) *Intertemporal Capital Asset Pricing Model* (ICAPM), which finds that the expected excess return of a risky asset may be different from zero even if the asset bears no systematic or market risk. However, the ICAPM is also not able to fully explain the empirical discrepancies found in the Black, Jenson, and Scholes study (Merton, 1973, p. 885) .

Another piece of evidence that the CAPM does not accurately describe stock returns is the fact that it is possible to forecast stock returns based on scaled price ratios, such as price-earnings or price-dividend ratios. According to the CAPM, only the systematic risk of a stock determines its expected return and hence, a forecast based on other factors should not be working. Campbell and Shiller (1988) show in their study that stock prices can indeed be predicted over the long-term using price-to-earnings ratios (p. 675) thus rejecting the CAPM hypothesis.

To summarize: The CAPM is a powerful tool to describe stock returns, but multiple studies have proven that it fails to explain several anomalies which result in statistically significant abnormal returns. In order to increase the explanatory power of asset pricing models, more factors are needed to account for the various effects which cannot be explained by a stock's beta. Multifactor models have been trying to overcome this problem with some success.

## 2.2 Multifactor Models

As discussed above, the CAPM as single-factor model fails to completely explain stock returns. To overcome this problem, multifactor models have been developed that use additional factors besides the market premium to explain stock returns. Consequently, an asset also has more than one factor loading, depending on the specific number of factors used.

A general theoretical framework for these multifactor approaches is the *Arbitrage Theory of Capital Asset Pricing* (APT) which was proposed by Ross (1976). Ross proposed that the expected return of a specific asset depends on a multitude of risk factors other than the single risk factor used by the CAPM (Ross, 1976, p. 355).

$$E_M = \sum_i \xi_i E_i$$

Where $E_M$ is the expected return, $\xi_i$ is the asset's sensitivity to the specific risk factor, and $E_i$ is the expected risk premium that arises from bearing that specific risk.

Chen, Roll, and Ross (1986) propose a set of macroeconomic variables that are to be used as risk factors in such a multivariate arbitrage pricing model. Examples for such risk factors would be the industrial production, unexpected inflation, or the term structure of interest rates. The complete set of variables examined in the Chen, Roll, and Ross paper can be seen in table 13 in the appendix. Due to their high explanatory power, macroeconomic variables are also used in other studies, such as in an attempt to describe the stochastic structure of asset prices (Cox, Ingersoll, & Ross, 1985, p. 383).

Another approach besides using macroeconomic factors to explain stock returns is the usage of variables which can be directly derived from stock returns or fundamental data. For example, an empirical study by Fama and French (1992) has examined the effects of a company's size, its book-to-market equity ratio, and its earnings-to-price ratio in order to build a multifactor asset pricing model that uses these factors as explanatory variables. Fama and French started with the classical CAPM equation, which is shown below:[2]

$$R_i - R_f = \alpha_i + b_i(RMRF) + \varepsilon_i$$

As previously discussed, there is an abundance of evidence that the CAPM ignores other risk factors affecting stock returns, such as the just discussed value and size effects of companies. This results in a relatively low explanatory power of this regression. As shown in table 14 in the appendix, the $R^2$ of this regression is relatively low with values around 0.80. In order to account for the effects of size and value on stock returns, Fama and French expanded the model and included two additional stock market factors that control for size and value effects (Fama & French, 1993, pp. 24-25). This three factor equation is shown below[3]:

$$R_i - R_f = \alpha_i + b_i(RMRF) + s_i(SMB) + h_i(HML) + \varepsilon_i$$

The SMB and HML factors are generated by sorting the whole universe of stocks into portfolios based on their size or market capitalization and their equity book-to-market ratios. Then, the return of the least risky portfolio is subtracted from the return of the most risky portfolio. The resulting difference in returns is the risk premium that enters the regression equation. As shown in table 15 in the appendix, both newly added factors have statistically significant factor loadings with most t-statistics larger than 2.0 and also the $R^2$ of the regression increases to values mostly about 0.90.

[2] RMRF = Excess market return
[3] SMB = Small minus big
HML = High minus low

In order to evaluate if stock returns can also be explained by bond market factors, Fama and French use a term structure premium as well as a credit risk premium as independent variables in a two-factor model to explain stock returns. The equation is shown below[4]:

$$R_i - R_f = \alpha_i + m_i(TERM) + d_i(DEF) + \varepsilon_i$$

The term structure premium is calculated as the difference in yield between long-term government bonds and the treasury-bill rate. The default risk premium is the difference between CB, which is a market proxy for average corporate bond yields, and the long-term government bond yield. Economically, this could be understood as the average credit spread of the market portfolio of debt.

The factor loadings and the corresponding t-statistics of this regression are shown in table 16 in the appendix. As shown in the table, the factor loadings are all non-zero and statistically highly significant with some t-statistics beyond 9.0. However, the explanatory power of this regression seems to be quite small with $R^2$ values below 0.20. Nonetheless, it is still notable that bond market factors explain at least some of the variation in stock returns. This result is also supported by the study of Chen, Roll, and Ross who find that both, the TERM as well as the DEF[5] factor have some explanatory power (Chen, Roll, & Ross, 1986, p. 402).

Fama and French consequently introduce a five factor model that uses both, stock and bond market factors, as variables to explain stock returns. The resulting equation is shown below (Fama & French, 1993, p. 28):

$$R_i - R_f = \alpha_i + b_i(RMRF) + s_i(SMB) + h_i(HML) + m_i(TERM) + d_i(DEF) + \varepsilon_i$$

---

[4] TERM = Term structure premium
DEF = Credit risk premium

[5] Chen, Roll, and Ross calculate the DEF factor as difference in yield between Baa-rated bonds and long-term government bonds. This is slightly different to the Fama-French methodology.

The risk premia which are used as factors in this regression are determined as previously discussed. The results of this regression can be seen in table 17 in the appendix. In this regression that includes stock as well as bond market factors the statistical significance of the factor loadings on TERM and DEF largely disappears with t-statistics mostly below 2.0. The significance of the RMRF, SMB, and HML factors however persists in this equation with some t-statistics beyond 50. Fama and French conclude that the statistical insignificance of TERM and DEF in this regression is due to the fact that RMRF captures most of the effects associated in these factors (Fama & French, 1993, p. 52). The fact that the TERM factor has a positive correlation of 0.37 to RMRF underlines this interpretation (Fama & French, 1993, p. 14). Also the $R^2$ of this regression increases only slightly compared to the previous three factor equation.

Another possible return explaining factor was identified by Jegadeesh and Titman (1993) in their study about momentum effects. Jegadeesh and Titman find that momentum portfolios, which buy stocks that gained in value during the last 12 months and short-sell stocks that lost in value during the last 12 months, generate abnormal returns not explained by systematic risk, size, or value factors. For example, a strategy which select stocks based on their past 6 month returns and holds them for 6 months realizes an average compounded excess return of 12.01% per year (Jegadeesh & Titman, 1993, p. 89). DeBondt and Thaler (1985) attribute this effect to substantial violations of weak market efficiency due to overreactions by investors to unexpected and dramatic events. DeLong, Shleifer, Summers, and Waldman (1990) have expended this view by adding that some investors knowingly try to front-run momentum effects therefore amplifying them creating a self-fulfilling prophecy. Hence, even though there might be no rational explanation for momentum effects, it is evident that these effects do exist in capital markets and should therefore be included in multifactor models trying to explain stock returns.

Carhart (1997) expands the Fama-French three factor model by adding such a factor that accounts for momentum effects in a study on mutual fund returns. The factor, PR1YR[6], is constructed as “the equal-weight average of firms with the highest 30 percent eleven-month returns lagged one month minus the equal-weight average of firms with the lowest 30 percent eleven-month returns lagged one month” (Carhart, 1997, p. 61). The intuition of excluding the last month's return in this analysis is to make sure that no short-term reversals are included in this factor. The resulting multifactor regression equation is shown below (Carhart, 1997, p. 61):

$$R_i = \alpha_i + b_i(RMRF) + s_i(SMB) + h_i(HML) + p_i(PR1YR) + \varepsilon_i$$

[6] PR1YR = Previous 1 year performance

This model works well with statistically significant factor loadings on all risk factors and is thus a useful extension to the traditional three factor model proposed by Fama and French (1993). The summary statistics of this four-factor model are shown in table 18 in the appendix. The resulting factor loadings and the adjusted $R^2$ are depicted in table 19 in the appendix. The factor loadings are statistically highly significant for most factors in most portfolios. Also the adjusted $R^2$ values are very high with values mostly beyond 0.90. Another interesting finding of Carhart's study is that when regressing mutual fund returns against this four-factor model, most funds systematically underperform and generate negative alphas which are statistically significant with some t-statistics beyond 3.0 (Carhart, 1997, p. 77). This result indicates that instead of hiring a fund manager, an investor would be better off by buying mimicking portfolios that track the four risk factors. This result becomes even more significant when accounting for fund management costs and transaction fees.

To conclude: The above examples as well as many other studies have revealed that multifactor models are a good method to explain stock returns and their causing factors. Multifactor models help to resolve anomalies not explained by the CAPM and have consistently higher $R^2$ than the classic CAPM regressions. Specific attention should however be paid to selecting the right factors and drawing the right conclusion from a specific factor. The danger of omitted variable bias can easily lead to statistical significant factor loadings although the true causal effect might be different. This danger is especially high if risk factors are highly correlated to each other.

## 2.3 The Advantages of Derivative Markets for Asset Pricing Models

In the last decade, derivative markets have grown significantly in size, product variety, and liquidity. Within some market segments, the notional value of outstanding derivatives is almost as large as or sometimes even larger than the face value of the underlying cash securities. Especially the credit default swap market has grown tremendously and is now one of the biggest and most liquid derivative markets. Recent studies estimate the notional value of outstanding CDS contracts as of December 2008 at over USD 41 trillion (European Central Bank, 2009, p. 4), other studies estimate an even higher notional value of about USD 58 trillion (US Securities and Exchange Commission, 2008).

This development and the growing importance of derivative markets for private as well as for institutional investors have generated new opportunities to trade various kinds of risks. The growing popularity of derivative markets can be attributed to an especially useful characteristic of financial derivatives. Whereas a cash market instruments usually bears a bundle of different risks, financial derivatives allow for the separate trading of these various risk kinds. For example, a traditional cash market bond is usually exposed to credit, interest rate, and currency risk. In the derivative market, all these kinds of risks can be traded separately from each other through the use of credit default swaps, interest rate swaps, and currency swaps or options. All of these products have the pleasant characteristic that they are virtually uncorrelated to the other risk types. For example, a credit default swap has almost no interest rate sensitivity. This separation of risk is especially useful for hedging applications. An investor who wants to hedge against rising interest rates can simply do so by entering an interest rate swap agreement. Using traditional cash market bonds, he would have to buy a floating rate note and finance this by issuing a fixed coupon bond. The payoff of the interest rate swap agreement and the cash market replication strategy are identical, but the complexity and transaction costs of the derivative transaction are much lower.

This separated trading of different risk types should also allow for interesting extensions of the previous discussed multifactor models. Since the returns of derivatives depend on risks which are almost uncorrelated to each other, introducing a multifactor risk model based on derivatives could improve forecasting power and minimize omitted variable bias. Since omitted variable bias occurs when an included factor is correlated to an omitted factor, the use of uncorrelated factors would help to overcome this problem. Also the traditional problem of not being able to distinguish between correlation and causality could be reduced since derivatives, at least to a large extent, only derive their value from only one specific kind of risk.

The faster absorption of news events in derivative markets is another advantage of using derivative implied factors as risk proxies in asset pricing models. An example for such a shorter reaction period is the lead-lag relationship of index futures to their respective indexes, where future price movements lead index movements by several minutes (Kawaller, Koch, & Koch, 1987, p. 1309). A similar argument could be constructed for the CDS market, which tends to react faster to news than the bond cash market (Daniels & Jensen, 2005, p. 31). Furthermore, derivative markets sometimes do not only react faster to certain events than cash markets, but they also price the same kind of risk differently than cash markets.

An empirical indicator for such a divergence of risk pricing between derivative and cash markets is the frequent possibility of negative default swap basis trades. The default swap basis is defined as the difference between the CDS spread and the cash LIBOR spread of a fixed income security, such as a fixed rate bond or a floating rate note (Fabozzi, 2005, p. 1347):

$$Default\ Swap\ Base = CDS\ Spread - Cash\ LIBOR\ Spread$$

According to classic economic theory, the default swap basis should be close to zero or positive to prevent arbitrage opportunities. Bearing the credit risk in form of a long position in a bond should yield the same return as bearing it in the form of selling a CDS. This also implies that holding a long position in a bond and insuring this position against default using a CDS should never yield more than the risk-free rate as return. If the CDS spread traded below the cash LIBOR spread, buying a bond in the cash market and insuring it against default would yield more than risk-free rate and thus result in an arbitrage opportunity. An example for such a trade is given below.

Assume a trader, who is able to refinance at LIBOR, has identified a floating rate note (FRN) paying LIBOR + 80bp p.a. and a counterparty willing to enter a CDS agreement with him at an annual CDS premium of 50bp. In this case, the trader could buy the bond and reinsure it against default and thus convert his long credit exposure into a virtually risk-free position[7]. Theoretically, he should earn the risk-free rate on this position, but the negative default swap basis allows him to make 30bp profit on this zero-cost portfolio. Clearly, this arbitrage opportunity should not exist in perfectly efficient capital markets.

[7] An additional assumption of this example is that the CDS counterparty is posting sufficient collateral so that counterparty risk can be neglected. This trade would, however, still not be perfectly risk-free since the trader is still exposed to the annuity risk that the bond issuer defaults before maturity.

**Picture 3: Negative Swap Basis Trade**

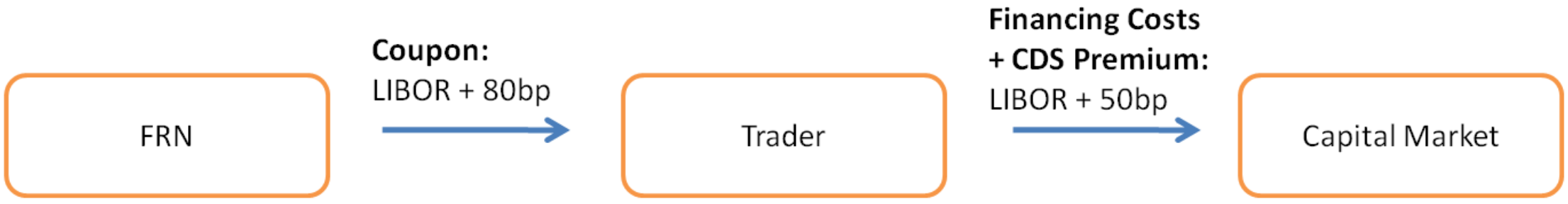

Such trades are widely known as *negative basis trades.* A negative basis is an arbitrage opportunity under the assumption that the trader has sufficient capital to maintain his positions when the swap basis temporarily widens. If cash and derivative markets moved synchronal, the swap basis should never be substantially different from zero. Hence, the abundance of negative basis opportunities is an indicator that cash and derivative markets are not perfectly correlated and sometimes price the same kind of risk differently.

Possible explanations for the divergent reaction of derivative and cash market to price relevant events are institutional externalities and short-selling constraints. For example, a long-only bond fund specializing in Greek sovereign bonds is expected to provide a positive exposure to Greek government bonds even if the fund manager regards the Greek government as not credit-worthy. In order to overcome this problem, the fund manager could enter a CDS agreement instead of selling bonds in the cash market. In this example, the additional demand would result in an increase in the CDS premium, but the absence of activity in the cash bond market would leave the credit spread unchanged. If the swap basis was previously at zero, it would now be positive.

The second possible explanation for divergent derivative and cash market spreads are short-sale constraints. Due to the higher regulation of cash markets, short-sale constraints are usually higher in these markets. Since short-sale constraints can lead to artificially inflated asset prices (Jones & Lamont, 2002), cash markets are more vulnerable to asset bubbles than derivative markets.

To conclude: Derivative markets offer some unique characteristics that make them promising alternatives to cash markets in asset pricing models. As discussed, derivative markets tend to react faster than cash markets, which are often restricted by institutional constraints. Sometimes derivative markets do not only react faster to events, but they also price risk differently than cash markets. The possibility of negative swap basis trades in the credit market is an example for such a divergent movement of markets. Perhaps the most significant advantage of derivative markets is the possibility of separating risk into smaller tradable risk types, which have only low sensitivities to each other. This is especially useful for multifactor models that require risk factors to be uncorrelated to each other. However, the fact that many derivative products are only traded in intransparent over-the-counter (OTC) markets is a disadvantage that should be kept in mind when using derivatives in asset pricing models. Sometimes it can be very difficult to gather the relevant information and even then it can be subject to massive biases.

# 3 Implied Volatility and CDS Spreads as Risk Factors

As a consequence of the rapid development of derivative markets, derivative instruments on almost all large-cap stocks and bonds are now available for trading. Two of the most important derivatives, according to their liquidity and trading volume, are equity options and credit default swaps. This paper utilizes these two products as base products for a multifactor stock pricing model. The model uses equity options to determine the implied volatility of a stock and the CDS spread to assess a company's credit riskiness or probability of default. Using these two products has some advantages: First, equity options as well as CDS are widely available for a large range of companies and are usually liquidly traded. The second advantage is that these products are standardized contracts[8], which eliminates the necessity of pre-processing large amounts of data before using them in an asset pricing model.

## 3.1 Implied Volatility

### 3.1.1 The Black-Scholes Formula

A standard model to calculate the implied volatility through equity option prices is based on the theories by Fischer Black and Myron Scholes (1973) as well as by Robert C. Merton (1973). Consequently, this model is widely referred to as the *Merton-Black-Scholes model*, for which the two then living authors Merton and Scholes received the Nobel Prize for economics in 1997. The Black-Scholes formula determines the price of a non-dividend paying European call option ($C$) based on its time to maturity ($T$), the spot price of the underlying asset ($S_0$), the option's strike price ($K$), the risk-free interest rate ($r$), and the volatility ($\sigma$) of the underlying asset (Hull, 2009, p. 291):

[8] Technically, only equity options traded on stock exchanges such as the Eurex or the CBOT are standardized contracts. CDS, which are not traded on stock exchanges, are not necessarily standardized. However, many market participants use the standardized master agreement proposed by the International Swaps and Derivatives Organization (2010).

$$C = S_0 N(d_1) - Ke^{-rt} N(d_2)$$

Where

$$N(x) = \Phi(x) = \ cdf\ of\ a\ standardized\ normal\ distribution$$

$$d_1 = \frac{\ln(\mathrm{S_0/K}) + \left(\mathrm{r} + \frac{\sigma^2}{2}\right) T}{\sigma\sqrt{\mathrm{T}}}$$

$$d_2 = \frac{\ln(\mathrm{S_0/K}) + \left(\mathrm{r} - \frac{\sigma^2}{2}\right) T}{\sigma\sqrt{\mathrm{T}}} = d_1 - \sigma\sqrt{\mathrm{T}}$$

The put price can be estimated through a similar function or through the application of the put-call parity, which states that the price of a put option can be derived from the price of a call option with identical strike price and maturity or (Hull, 2009, p. 208):

$$c + Ke^{-rt} = p + S_0$$

Economically, this means that buying a stock and hedging it against losses with a put option should have the same expected payoff as buying a call option and investing the remaining capital in the risk-free asset.

**3.1.2 Determining the Implied Volatility**

All parameters of the Black-Scholes pricing formula are known in advance, except the volatility since it is not the historical volatility which is relevant for the price determination, but the expected future volatility. The volatility which is implied by the option price is hence known as the *implied volatility* (Hull, 2009, p. 296). In order to determine this implied volatility, traders accept the price of an option as given and then back-solve for the implied volatility that sets equal both sides of the Black-Scholes equation. In practice however, algebraic back-solving is not sufficient to determine the implied volatility of more complex options, such as American options. Also the assumptions of the Black-Scholes model, such as a constant risk-free interest rate, constant implied volatility, or absence of dividend payments make it necessary for traders to use more complex approaches to determine option prices or implied volatilities, such as the *implied volatility function* (IVF) model developed by Derman and Kani (1994), Dupire (1994), and Rubinstein (1994).

For the scope of this paper however, it is sufficient to use implied volatility data that is calculated under the simplifying assumptions of the Black-Scholes formula. For further future studies, more complex implied volatility determination methods could theoretically be applied in order to gain more exact and accurate results.

### 3.1.3 Implied Volatility and Stock Returns

In order to be a useful factor to explaining stock returns, implied volatility should have significant effects on stock returns. Recalling the classic positive relationship between risk and return, it seems obvious that higher expected risk would result in higher expected returns:

$$E(\sigma_t) \sim E(r_t)$$

Yet, empirical results trying to test this relationship are somewhat inconclusive. Cambell and Hentschel (1992) and French, Schwert, and Stambaugh (1986) find evidence for a positive correlation between expected returns and expected risk, whereas Glosten, Jagannathan, and Runkle (1993) find a negative relationship between these two factors.

Due to the fact that a single asset always bears some non-systematic risk to a certain extent, both findings do not necessarily oppose classic economic theories which state that only systematic risk is being rewarded.

$$Var[\tilde{r}_i] = \beta_{iM}^2 \, Var[\tilde{r}_m] + Var[\tilde{\epsilon}_i]$$

Due to the fact that an expected increase in volatility cannot ad-hoc be appointed to the systematic or the non-systematic component of total risk, no prediction based on implied volatility about the development of future returns can be made under the assumption that only systematic risk is being rewarded. The effect on the stock’s return would depend on how much of the expected risk can be allocated to the systematic component of the total risk. If one was able to forecast only the systematic portion of the overall risk, return forecasts based on this systematic risk should theoretically be possible. Nonetheless, since markets are not perfectly efficient, any a-priori statement about the development of future returns would still be difficult. The regression analysis later in this paper will hopefully generate some insights about the relationship between implied volatility and stock returns and about the possibility of using implied volatilities as asset pricing factor.

## 3.2 The CDS Spread and Stock Returns

The second factor which is used in this paper to explain stock returns is the credit risk of the specific company. Specifically, the CDS spread of the company over LIBOR is used as proxy to determine the credit risk. Numerous papers have identified factors, such as the dividend payout ratio (Lamont, 1998), which are correlated to stock returns as well as bond yields. These findings are an implication that stock and bond returns are either directly correlated to each other or correlated to each other through external factors. If that was true, bond market factors could indeed be helpful in explaining stock returns.

Fama and French (Fama & French, 1989, p. 48) find that the default premium on bonds can help to explain stock returns due to the fact that it is correlated with other relevant factors, such as the dividend yield or the overall business cycle. It is likely that a credit risk factor is also highly correlated to other commonly used factors to explain stock returns. Especially the HML factor of the Fama-French model, which is a proxy for the market valuation of a company's equity, is a candidate for such a relationship since it can be seen as a distressed factor (Fama & French, 1998) which should also be picked up by higher default expectations in the bond market. Newer studies go even further in examining the relationship between the stock and the bond market and try to explain CDS spreads using equity options (Zhang, Zhou, & Zhu, 2009). Also the development of joint frameworks for the valuation and estimation of stock options and CDS (Carr & Wu, 2009) indicate a close relationship of stock and bond markets. Norden and Weber (2009) have not only found a strong relationship between these markets, but also found that stock returns are significantly negatively correlated with CDS and bond yield changes (p. 554). However, it remains to be seen if those results, which origin from a small sample and a very limited observation period, apply to more general test settings.

All these theories and studies have shown significant evidence that relationship patterns exist between the stock and the bond market. This paper will therefore use the CDS spread as risk factor to account for the market's perception of the credit worthiness of the specific company. If credit worthiness affected the stock price development, it should be reflected in a statistically significant factor loading on the CDS factor in the upcoming multivariate regression model.

# 4 Impact of Credit Risk and Implied Volatility on Stock Returns

## 4.1 Credit Risk and Implied Volatility as Risk Factors

As discussed before, this paper tries to explain stock returns through factors based on the CDS spread of a company and its equity option implied volatility. Previous theories have already analyzed the effect of credit spreads on stock returns and found a positive relationship between credit risk and stock returns. The following section further advances on this topic by using credit default swap spreads as input to calculate a CDS spread risk premium. The corresponding factor loadings are determined through a multivariate regression analysis. If successful and in line with previous studies, the regression model developed in this paper should also pick up a positive return premium for stocks with higher credit risk. Special emphasis in this paper is also paid to the distribution of the factor loadings according to a stocks market-to-book ratio, which is a proxy for a stock's *value-effect* as described by Fama and French (1993) and others. The effect of implied volatility on stock returns has also been studied before with ambivalent results. Some studies found positive effects of higher implied volatility on stock returns whilst other studies have concluded that there is no measurable effect or even a negative effect. In order to resolve this problem, this paper first determines the risk premium of higher implied volatility and then estimates factor loadings on this factor.

## 4.2 Analytic Methodology

### 4.2.1 Stock Universe

The core universe used in this paper consists of the stocks included in the Standard & Poor's 100 Index, commonly referred to as the S&P 100. The intention of using this index is that the stocks of its 100 constituent companies are very liquidly traded and that equity as well as credit derivatives are easily available. However, credit default swap data with sufficient quality is only available for 83 of these 100 companies; hence the universe has to be reduced to these 83 companies. Please see table 20 in the appendix for the complete list of stocks and the corresponding credit default swaps which were used for this paper.

When available, standard CDS contracts with a five year maturity are used. However, for some companies only CDS with different maturities are actively traded. In these cases, the CDS with the closest maturity to five years are used instead. Equity options are easily available for all companies in the S&P 100 Index. In order to gather the implied volatility, the implied volatility of at the money call options is used[9], thereby ignoring possible relationships between different strike prices and different implied volatilities, known as volatility smiles. Nonetheless, for the purpose of this paper and for the sake of simplicity, the assumption of constant implied volatilities across different strike prices should have no significant effect on the research results[10].

The data for the implied volatilities is only available from May 15th 2008 to January 15th 2009, therefore drastically reducing the external validity of any results since only a very small timeframe can be analyzed. The data for stock prices and CDS spreads is easily available over a five year time horizon, which is used for any analyses that do not include factors based on the implied volatility.

#### 4.2.2 Econometric Methodology

The main econometrical method used in this paper is the multivariate regression analysis. Where appropriate, the sample is broken down into quintiles or deciles to account for possible differences in factor loadings across different portfolios. In all regression analyses, HAC standard errors are used to control for heteroskedasticity and autocorrelation.

The general approach of building the regression model used in this paper is to first calculate the risk premia for stocks with higher credit default swap spreads and higher implied volatilities. These risk premia are determined by subtracting the equal weighted average return of stocks with the lowest CDS premia or implied volatility from those stocks with the highest CDS and implied volatility risk premia.

$$Risk\ Premium = Return_{High-risk\ stocks} - Return_{Low-risk\ stocks}$$

The second step in building the regression model is to create quantile portfolios of stocks against which the risk premia are regressed. Since this paper tries to examine effects which are probably highly correlated to the market-to-book (MB) ratio of a company's equity, the portfolios are created with the MB ratio as sorting criteria. The stocks with the lowest MB ratio enter portfolio one, the stocks with the next lower MB ratio enter portfolio two, and so forth.

[9] The implied volatility used in this paper is the implied volatility which is supplied by the DataStream Financial Database, which uses a standard Black-Scholes model for its determination. This method uses call options with at or near the money strike prices.

[10] An effect that might be significant is the lack of transparency in the pricing of CDS contracts. Since CDS are OTC traded products, it is possible that the quoted spreads by market makers, which are used in this paper, do not appropriately reflect the true spreads agreed on in large OTC transactions.

Having created these quantile portfolios sorted by the MB ratio, their equal weighted average return is regressed against the risk factors, which are the risk premia that are determined as described above. Furthermore, the market return is included in some of the regression to control for overall market developments which affect all stocks in the universe. The resulting factor loadings, their statistical significance, and the explanatory power of the regression are then discussed.

## 4.3 Determination of Risk Premia

### 4.3.1 The Risk Premium for CDS Spreads

In order to determine the risk premium for credit risk, on each day the universe of stocks is sorted according to the companies' CDS spreads. After sorting the companies according to their CDS spreads, the stocks of the companies with CDS spreads below the 25$^{th}$ percentile of all CDS spreads are selected to form the *unrisky* (U) portfolio. This procedure is repeated for the companies with CDS spreads above the 75$^{th}$ percentile to create the *risky* (R) portfolio. Afterwards, the equal weighted average returns of the stocks in the risky and the unrisky portfolio are calculated. The CDS premium factor is then calculated by subtracting the return of the low-risk portfolio (U) from the high-risk portfolio (R). The resulting portfolio, which is long the risky portfolio and short the unrisky portfolio, is called RMU which stands for *risky minus unrisky*. This portfolio is rolled over and regenerated every single trading day. The summary statistics for the daily returns of this RMU portfolio, as well as for the two constituent portfolios R and U are shown below:

**Table 1: Daily Summary Statistics for RMU and RM**

| | RMU | Risky (R) | Unrisky (U) | RM |
|---|---|---|---|---|
| **Mean Return** | 0.0002358 | 0.0003609 | 0.0001250 | 0.0003024 |
| **Standard Error** | 0.0004055 | 0.0006359 | 0.0003281 | 0.0004220 |
| **t-statistic** | 0.582 | 0.567 | 0.381 | 0.717 |
| **N** | 1304 | 1304 | 1304 | 1304 |

As shown in the table, the risky portfolio earns a daily return of roughly 0.0236% during the full five year observation period. This result is not statistically significant on a daily horizon with a t-statistics of 0.58. Also the daily return 0.036% of the risky portfolio and the daily return of 0.0125% of the unrisky portfolio are not statistically different from zero on a daily level. Nonetheless, this level of statistical significance is in line with that of the daily return of the market, which is calculated as the equal weighted average return of all securities in this paper's sample universe. This market portfolio (RM) has a return of 0.03024% on a daily level and is also statistically not significant with a t-statistic of 0.717.

Although the daily return of the RMU portfolio seems low at first, it implies an average excess return of the risky portfolio compared to the unrisky portfolio of around 5.89%[11] on an annual base. Also the low statistical significance of the daily returns can be reduced when transforming the summary statistics to an annual level[12] as shown in the table below:

**Table 2: Annual Summary Statistics for RMU and RM**

| | Daily | | Annually | |
|---|---|---|---|---|
| | RMU | RM | RMU | RM |
| **Mean Return** | 0.0002358 | 0.0003024 | 0.058962 | 0.075605 |
| **Standard Error** | 0.0004055 | 0.0004220 | 0.006411 | 0.006673 |
| **t-statistic** | 0.582 | 0.717 | 9.2*** | 11.3*** |

***p-value < 0.001

As can be seen in the table, transforming the daily summary statistics to an annual level results in t-statistics that correspond to confidence levels beyond 99.9% for both, the RMU as well as the RM portfolio. Consequently, it can be concluded that the RMU portfolio generates positive returns which are statistically different from zero on an annual level. The most risky stocks, as measured by their CDS spread, outperform the least risky stocks by 5.8% per year.

### 4.3.2 The Risk Premium for Implied Volatility

Similar to the approach just described to determine the RMU factor, the risk premium for the implied volatility is calculated. As discussed, data on the implied volatility is only available for a shorter observation period starting May 20th 2008. Hence, the number of observation days is reduced to 436 days.

[11] Under the assumption of 250 annual trading days.

[12] Transforming is conducted by multiplying the return with 250, the number of trading days per year, and the standard error with the square root of 250.

The risk premium portfolio for the implied volatility is called *VMC*, which stands for *volatile minus consistent*. As the name already tells, this factor is the difference of the equal weighted average return of stocks with high implied volatility and the equal weighted average return of stocks with low implied volatility. Specifically, the universe of stocks is sorted on a daily level by the individual stock's implied volatility. Then, the stocks with an implied volatility below the 5th percentile of all implied volatilities form the *consistent* portfolio (C), whereas the stocks with an implied volatility beyond the 95th percentile form the *volatile* portfolio (V). The VMC portfolio is then simply the difference in returns of these two portfolios. The summary statistics for the two portfolios and VMC are shown below.

**Table 3: Daily Summary Statistics for VMC**

| | **VMC** | **Volatile (V)** | **Consistent (C)** |
|---|---|---|---|
| **Mean Return** | -0.0031596 | -0.0019835 | 0.0011761 |
| **Standard Error** | 0.0024779 | 0.0028300 | 0.0007404 |
| **t-statistic** | -1.275 | -0.701 | 1.588 |
| **N** | 436 | 436 | 436 |

As shown, the stocks which have an implied volatility beyond the 95th percentile underperform the stocks with implied volatilities below the 5th percentile by 0.32% per day. This effect is statistically not significant on a daily level with a t-statistic of 1.275. When looking at the returns on an annual level, the underperformance of the volatile portfolio becomes remarkably large.

**Table 4: Annual Summary Statistics for VMC**

| | **VMC** | |
|---|---|---|
| | Daily | Annually |
| **Mean Return** | -0.0031596 | -0.7898949 |
| **Standard Error** | 0.0024779 | 0.0391792 |
| **t-statistic** | -1.3 | -20.2*** |

***p-value < 0.001

The summary statistics above state that the volatile stocks underperform the non-volatile stocks by almost 79% per year. This result is statistically significant on a 99.9% confidence level. This extremely large underperformance seems unrealistic at first glance and hence further analysis is needed to understand the origins of this remarkable result.

When looking at the time-series of VMC returns in more detail, it becomes obvious that the massive underperformance of the volatile stock portfolio is largely driven by the first 285 days in the sample. During this period, the volatile portfolio underperforms the consistent portfolio by 0.5% per day. Annualized, this is an underperformance of around 125%. In the subsequent 151 days, that trend changes and the volatile portfolio is being underperformed by 0.05% a day or 12.5% annually.

**Table 5: VMC Time-Series Split**

| | VMC | |
|---|---|---|
| | Until 06/22/09 | After 06/22/09 |
| **Mean Return** | -0.0051052 | 0.0005126 |
| **Standard Error** | 0.0036872 | 0.0016408 |
| **N** | 285 | 151 |

The most probable explanation for this structural break is a sample selection bias. With an overall number of observations of 436, the sample is very small and includes less than two years of data. Furthermore, a majority of the data in this sample was collected in the middle of a massive financial crisis and the external validity of any results based on this sample should be carefully questioned. The massive underperformance of riskier stocks shown above could very well be a misleading snapshot that was taken during a time when massive deleveraging was taking place and investors sold tremendous amounts of risky assets therefore depressing prices of risky assets. This period was also characterized by extremely high levels of volatility, as the chart below clearly shows.

**Picture 4: VMC Daily Returns**

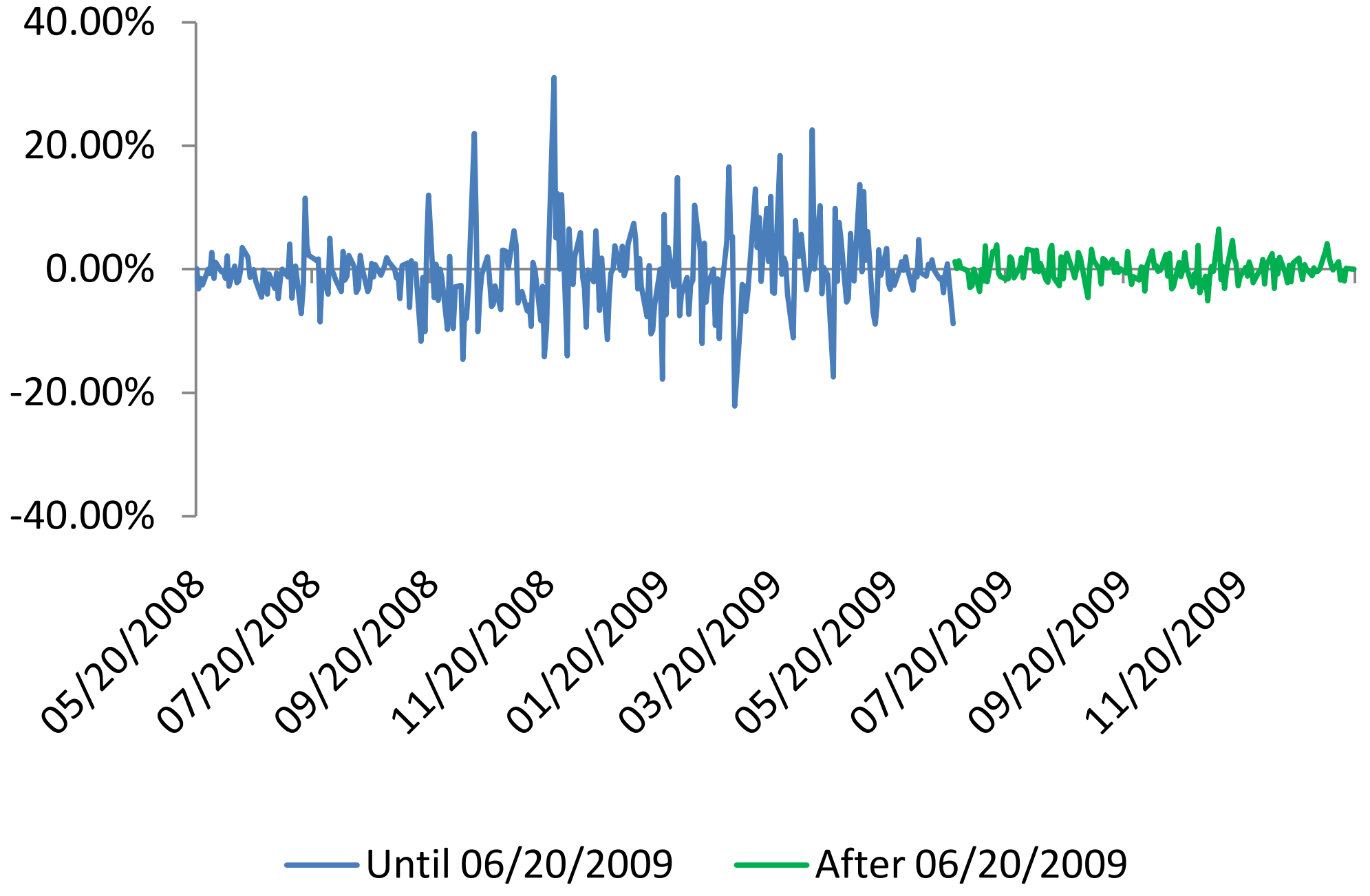

This high level of volatility ends abruptly in June 2009. This timing corresponds to the time, when the daily *underperformance* of risky assets transforms into a daily *outperformance* of risky assets, as shown above. The most likely explanation for this behavior is a change in risk appetite by many investors as a result of the financial crisis, which required investors to massively deleverage. As a consequence of this deleveraging and the overall reduction in investor's willingness to take risk, investors disposed their most risky assets thus depressing share prices of stocks with higher implied volatility. When this deleveraging and sellout of risk assets ended in June 2009, investors started to reposition themselves by increasing their overall exposure to risky assets. As investors increased their positions in these now low-priced assets, the share prices of these riskier stocks started increasing more than the share prices of stocks with lower implied volatility.

To conclude: The risk premium for implied volatility is extremely negative. This result is statistically highly significant on an annual level. However, this result is probably subject to major sample selection bias since the observation period mainly comprises of a time period which was characterized by dramatic breaks in market behavior, flights towards safer assets, and a massive economy-wide deleveraging. Hence, it is evident that more data, which includes at least one full economic cycle, would be very useful in increasing the validity of the results presented.

## 4.4 Regression Analysis

### 4.4.1 CDS Premium

In order to estimate the factor loadings for the CDS premium factor, the stocks are first sorted into five quintiles according to the market-to-book (MB) ratio of their equity. The companies with the lowest MB ratio enter quintile one, the stocks with the highest MB ratios enter quintile five. For each quintile, the equal weighted average return of the companies' stocks is calculated. This procedure is repeated every single trading day.

The basic intuition of sorting the stocks by their MB ratio is to control for the value effect of stock returns. The value effect states that stocks with lower MB ratios outperform stocks with higher MB ratios. Fama and French (1998) proposed that this value effect could actually be a consequence of the fact that companies with low MB ratios are often distressed companies. If this was true, the regression analysis should generate higher factor loadings on RMU for lower MB quintiles since distressed companies usually have higher credit spreads. The five MB quintiles are then regressed against the risk premium for credit default swap spreads, RMU.

$$R_P = \alpha_P + c_P RMU$$

**Table 6: MB Quintiles regressed against RMU**

| | MB Quin~1 | 2 | 3 | 4 | MB Quin~5 |
|---|---|---|---|---|---|
| rmu | 1.454*** | 0.816*** | 0.585*** | 0.535*** | 0.425*** |
| | (29.64) | (18.54) | (15.54) | (13.76) | (10.69) |
| _cons | -0.000273 | 0.000112 | 0.000141 | 0.000329 | 0.000341 |
| | (-0.85) | (0.34) | (0.47) | (1.10) | (1.23) |
| N | 1304 | 1304 | 1304 | 1304 | 1304 |
| R-sq | 0.771 | 0.504 | 0.384 | 0.343 | 0.278 |

t statistics in parentheses
* p<0.05, ** p<0.01, *** p<0.001

As expected, the factor loadings on RMU are larger for the lower MB quintiles. The factor loading on RMU for the first quintile is 3.4 times higher than the factor loading of MB quintile five. Both factor loadings are statistically significant at a 99.9% confidence level. This decrease in factor loadings supports the above mentioned theory that the value effect is at least partially a consequence of a credit risk premium earned by distressed companies' stocks. In all regressions, the constants are statistically insignificant.

The $R^2$ values also decrease for portfolios with higher MB ratios. The quintile one portfolio has an $R^2$ of above 0.77, whereas the fifth quintile portfolio has an $R^2$ of only 0.28. This decrease in the explanatory power of the credit risk premium factor RMU bears an interesting finding: It seems that the credit risk has a very high explanatory power for portfolios or stocks with high credit risk, but only a very modest explanatory power for stocks with lower credit risk. This is an indication that the credit risk of a specific company has a high influence on the stock price when the company is in financial distress, but only low influence on companies that are in financially stable condition. As soon as a company's financial situation improves, the credit risk is no longer an important determinant of stock returns. This seems perfectly logical due to legal and accounting reasons: Credit risk is only important for a shareholder when a company is in imminent danger of bankruptcy, in which case a shareholder usually loses his whole investment. If a company is financially stable, the stock return should theoretically be almost independent of the credit risk. This is very well reflected in the factor loadings and also the explanatory power; both decrease for companies with less credit risk.

In order to control for the overall market development, the equal weighted average return of all stocks in the sample universe (RM) is added to the regression of the five quintiles.

$$R_P = \alpha_P + c_P RMU + \beta_P RM$$

**Table 7: MB Quintiles regressed against RMU and RM**

| | MB Quin~1 | 2 | 3 | 4 | MB Quin~5 |
|---|---|---|---|---|---|
| rmu | 0.681*** | -0.0153 | -0.171*** | -0.231*** | -0.276*** |
| | (24.04) | (-0.70) | (-9.69) | (-16.66) | (-17.67) |
| rm | 1.010*** | 1.086*** | 0.988*** | 1.000*** | 0.916*** |
| | (49.33) | (54.19) | (50.00) | (74.07) | (54.05) |
| _cons | -0.000396** | -0.0000204 | 0.0000210 | 0.000207* | 0.000230* |
| | (-2.83) | (-0.19) | (0.21) | (2.30) | (2.50) |
| N | 1304 | 1304 | 1304 | 1304 | 1304 |
| R-sq | 0.956 | 0.948 | 0.930 | 0.941 | 0.920 |

t statistics in parentheses
* p<0.05, ** p<0.01, *** p<0.001

When including the equal weighted average returns of all stocks in this study's universe (RM), the factor loadings on RMU decrease but remain statistically significant. The $R^2$ values of the regression increase to values beyond 0.90, thus explaining most of the variation in stock returns.

As before, the factor loadings on RMU are much higher for the lower MB quintile stocks than for the high MB quintile stocks. When adding RM as factor to the regression, an interesting fact becomes evident: The factor loading on RMU is positive only for the first quintile portfolio but negative for all other quintiles. These loadings are statistically significant at a 99.9% confidence level, except for the second quintile portfolio. This is however due to the fact that the factor loadings turn from positive into negative and hence the factor is not statistically significant different from zero, which is what the t-statistic measures. This development of the RMU factor loadings, which turn from positive to negative, is even better visible when looking at the regression with a higher resolution by dividing the stock universe in deciles instead of quintiles. Please see table 21 in the appendix for the regression results of MB deciles against RMU and RM. The development of the factor loadings on RMU of this decile-regression is shown below:

**Picture 5: RMU Factor Loadings of Decile Regression**

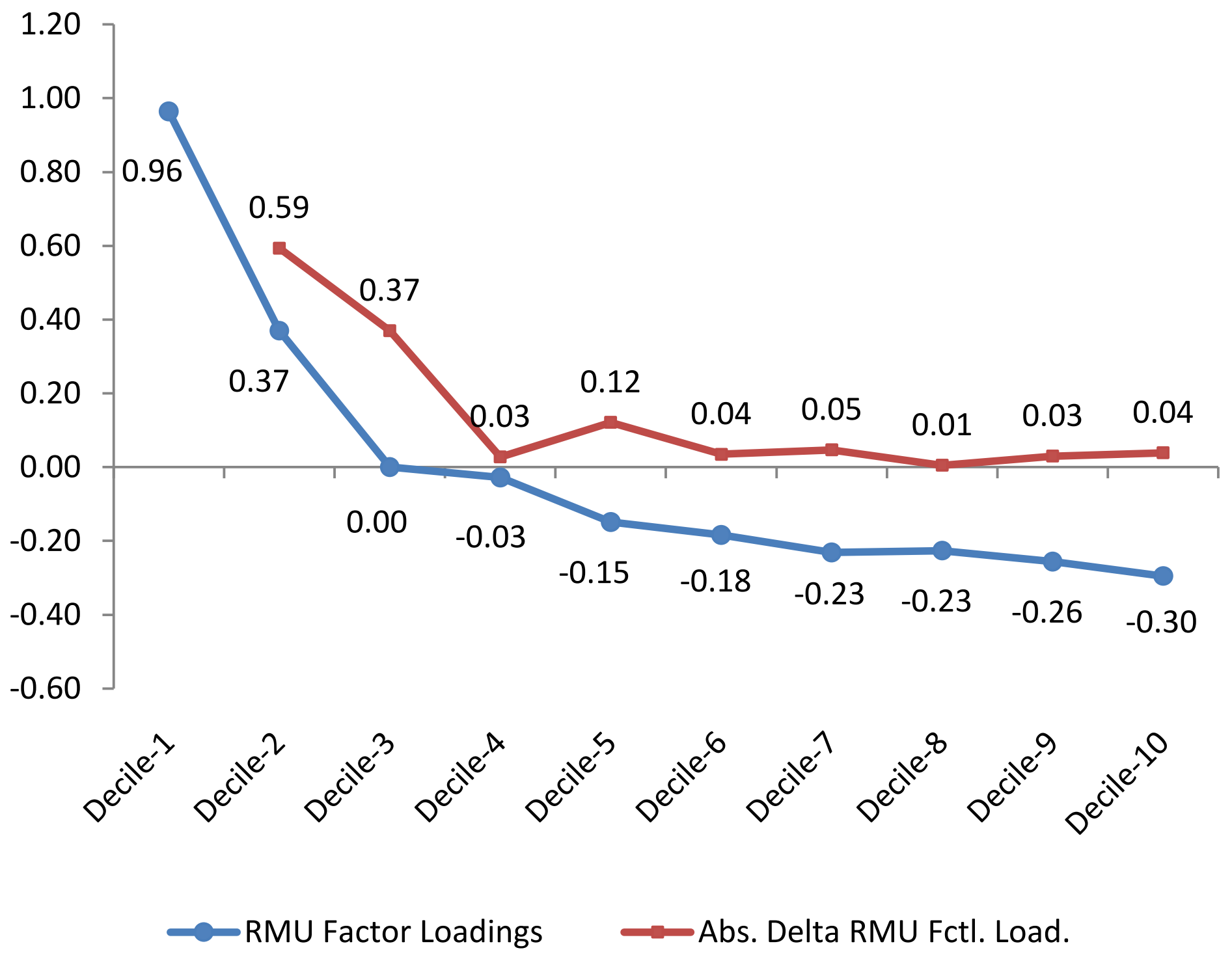

As already discovered in the quintile regression, the factor loadings on RMU decrease for portfolios that have higher MB ratios. The factor loadings are positive for the first two decile portfolios and eventually turn negative for the other portfolios. These factor loadings are statistically significant different from zero for most factor loadings except for the MB deciles in which the factor loadings turn from positive to negative and which are logically very close zero.

This development of the RMU factor loadings shows that stocks of companies with higher credit risk earn positive return premiums. The stocks of companies with lower credit risk earn negative return premiums. It seems that shareholders who bear a higher amount of default risk are compensated with a stock return premium for bearing that risk. On the other hand, shareholders who invest in safer stocks with lower credit risk are receiving a negative return premium. However, the positive return premium for companies with high credit risk is much higher than the negative return premium for companies with low credit risk. This is also reflected by a decrease in the absolute difference in the factor loadings when comparing low quantile portfolios to high quantile portfolios. The difference in factor loadings from the decile one to decile two is much higher than the difference from decile nine to decile ten. Graphically, this is reflected in the diminishing differential of the factor loadings between decile portfolios, depicted as red line in the above chart. This is an indication that as soon as the credit risk of a company falls below a certain threshold, it has no influence on stock returns anymore. In clear words: It does matter if a company's financial profile changes from high credit risk to medium credit risk, but changing from low credit risk to extremely low credit risk has almost no effect on stock returns. This is also underlined by the development of the $R^2$ values for the different quintiles in the regression analysis. The $R^2$ figures decrease for companies with higher MB ratios, implying that the RMU factor and thus credit risk is less helpful in explaining stock returns for companies with higher MB ratios or lower credit risk.

From the above presented results it becomes evident that CDS spreads, as proxies for credit risk, are helpful in explaining stock returns. Companies with higher CDS spreads tend to have higher stock returns than companies with lower credit risk. This relationship is stronger for distressed companies, which have low market to book valuation ratios. The stocks of those distressed companies are earning excess returns for bearing the risk of default. Stocks of companies with higher MB ratios and thus lower credit risk generate no premium returns; they even earn discounts as a result of the lower credit risk. This is shown by negative RMU factor loadings for the higher quantile portfolios when RM is also included in the regression analysis. The explanatory power of CDS spreads on stock returns decreases for higher market to book ratios, indicating that the stock returns of companies which are not distressed are less affected by credit market fluctuations. Once the danger of bankruptcy is not imminent anymore, shareholders worry about things other than the credit risk of the company. At this point, other factors such as the corporate strategy and the economic development are more likely to capture the investors' attention. This is reflected in declining factor loadings, t-statistics, and $R^2$ for companies with increasing MB ratios. Economically this implies that credit risk has huge effect on stocks returns only for companies in which the credit risk is very high. As soon as the credit risk has fallen below a certain level, the explanatory power of the credit risk premium RMU rapidly deteriorates. Another interesting finding of this regression is the statistically significant negative intercept for the quintile 1 portfolio regressed against RM and RMU. Even when controlling for market return and CDS risk premium, this portfolio generates negative excess return of roughly 9.9% per year. This intercept is statistically significant on a 99% confidence level.

### 4.4.2 Volatility Premium

For the regression estimation of the factor loadings on the volatility premium factor VMC, the sample size has to be decreased to 436 trading days since data on implied volatility is only available for a shorter time horizon. The time frame used for this analysis starts on May 20$^{th}$ 2008 and ends on January 19$^{th}$ 2010. As in the previous section with the CDS premium factor loadings, a quintile regression is conducted in which the equal weighted average returns of five portfolios sorted by their MB ratios are regressed against the VMC and the RM factor. The results of the regression against the VMC factor alone are shown below:

$$R_P = \alpha_P + v_P VMC$$

**Table 8: MB Quintiles regressed against VMC**

| | (1) MB Quin~1 | (2) 2 | (3) 3 | (4) 4 | (5) MB Quin~5 |
|---|---|---|---|---|---|
| vmc | 0.663*** | 0.363*** | 0.248*** | 0.236*** | 0.186*** |
| | (23.27) | (13.56) | (11.10) | (10.42) | (7.83) |
| _cons | 0.00210* | 0.00105 | 0.000836 | 0.00104 | 0.000835 |
| | (2.19) | (1.21) | (1.08) | (1.35) | (1.19) |
| N | 436 | 436 | 436 | 436 | 436 |
| R-sq | 0.752 | 0.518 | 0.390 | 0.365 | 0.313 |

t statistics in parentheses
* p<0.05, ** p<0.01, *** p<0.001

The factor loadings on VMC are all statistically significant on a 99.9% confidence level and decrease for portfolios with higher MB ratios. The $R^2$ values for the regressions on the different quintiles also decrease for portfolios with higher MB ratios. It should however be kept in mind that during this observation period the VMC return is negative, hence a positive factor loadings means that a stock is earning a negative return premium.

**Table 9: VMC Summary Statistics**

| Variable | Obs | Mean | Std. Dev. | Min | Max |
|---|---|---|---|---|---|
| vmc | 436 | -.0031596 | .0517403 | -.2216613 | .310394 |

Due to the short observation period which falls right into the middle of the financial crisis, these results might mainly be caused my massive market devaluations and not by the true causes this paper tries to identify. To minimize this problem the equal weighted average return of the market portfolio RM is added to the regression to control for market developments.

$$R_P = \alpha_P + v_P VMC + \beta_P RM$$

**Table 10: MB Quintiles regressed against VMC and RM**

| | (1) MB Quin~1 | (2) 2 | (3) 3 | (4) 4 | (5) MB Quin~5 |
|---|---|---|---|---|---|
| vmc | 0.279*** | -0.00692 | -0.0808*** | -0.0961*** | -0.102*** |
| | (18.40) | (-0.57) | (-9.89) | (-11.74) | (-11.78) |
| rm | 1.131*** | 1.089*** | 0.968*** | 0.977*** | 0.849*** |
| | (30.03) | (42.53) | (45.06) | (49.59) | (35.26) |
| _cons | 0.000776 | -0.000225 | -0.000297 | -0.0000998 | -0.000159 |
| | (1.96) | (-0.86) | (-1.31) | (-0.50) | (-0.67) |
| N | 435 | 435 | 435 | 435 | 435 |
| R-sq | 0.957 | 0.958 | 0.948 | 0.956 | 0.923 |

t statistics in parentheses
* p<0.05, ** p<0.01, *** p<0.001

The results show that the size of the VMC factor loadings decrease as soon as the market return is added to the regression as additional factor. The VMC factor loadings are still all significant at a 99.9% confidence level, except for the factor loading of MB quintile 2. The insignificance of this factor is true to a downward trend in VMC factor loadings for increasing MB ratios which turn from positive to negative. Economically this means that portfolios with larger MB ratios are earning a premium for being less volatile. On the other hand, stocks in the low MB portfolio have a positive factor loading on VMC which indicates that they are more volatile and consequently generate negative stock return premiums. Again, it shall be noted that this effect, although being statistically highly significant, might only be a result of the observed time period which was characterized by a massive global economic downturn and panic sales of risky assets. In order to gain a better understanding of this factor, further studies need to be conducted with an observation period which includes at least one full economic cycle. Nonetheless, a graph of the resulting factor loadings on VMC of a regression of the MB deciles against VMC and RM is shown below. For the complete table of regression results please see table 21 in the appendix.

**Picture 6: VMC Factor Loadings of Decile Regression**

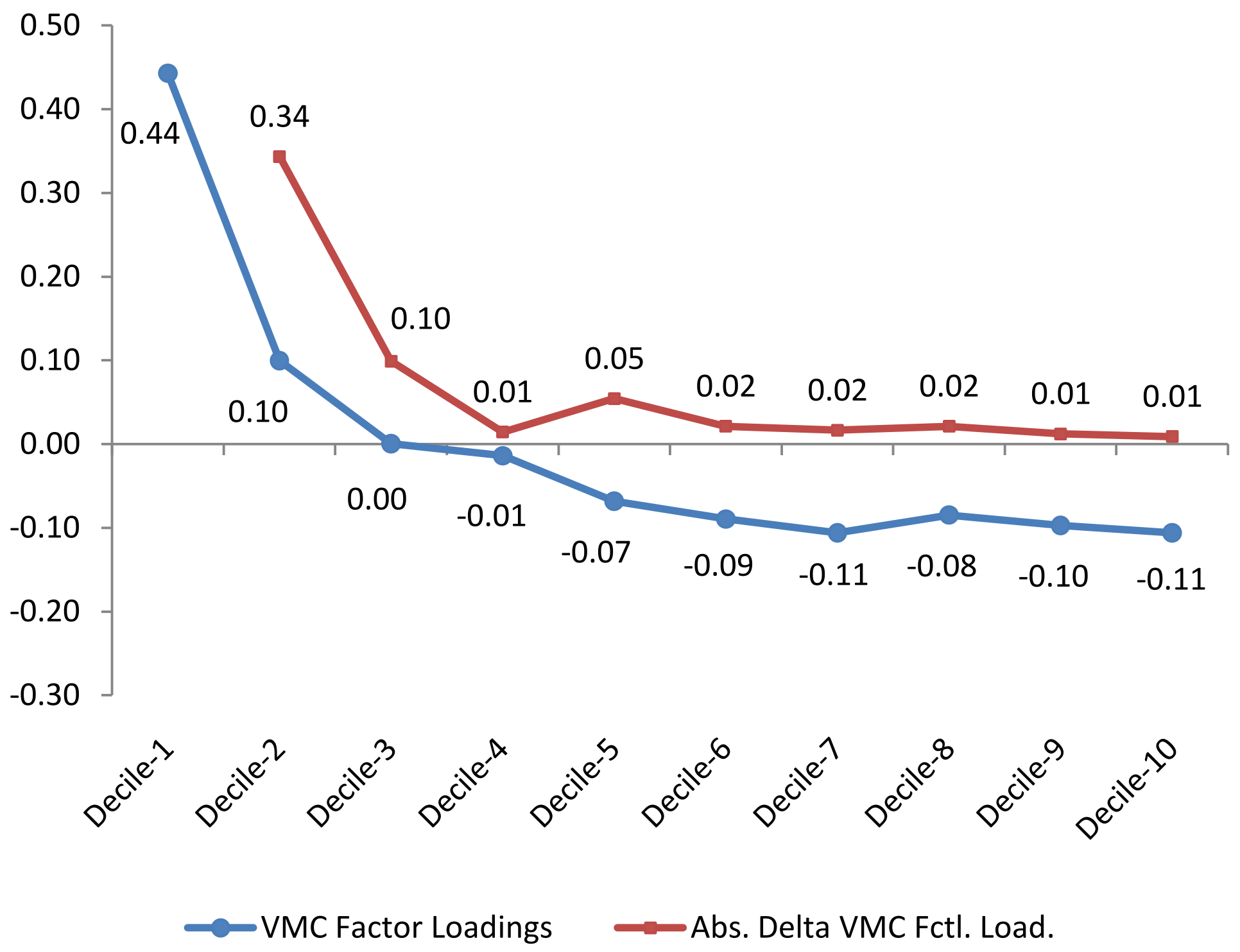

As shown in the graph, the factor loadings are positive for portfolios with low MB ratios and turn negative for portfolios with higher MB ratios. The rate of change however rapidly deteriorates as soon as the MB ratios become higher and is approximately constant for quintiles six to ten. Economically this indicates that stocks with low MB ratios, which can be seen as distressed stocks, have excess implied volatility and have consequently earned massive negative premiums in the observed time period. Stocks with higher MB ratios tended to be less volatile and thus generated premiums for being less risky assets. However, the premium earned by less volatile stocks is much smaller than the negative premiums earned by the volatile stocks in the lowest MB decile.

To conclude this section: Stocks with low MB ratios tend to be more volatile in the observed time period. During this time period, volatility is highly correlated with negative returns. Hence, stocks with low MB ratios have underperformed those with higher MB valuations. As previously discussed, this outcome should be subject to careful questioning since the observation period is very short and falls into the middle of the financial crisis. Therefore the result just presented could be biased and its external validity should be subject of further studies.

### 4.4.3 CDS-, Volatility-, and Market Premium

As previously discussed, it is evident that the CDS premium factor and the implied volatility premium factor are helpful in explaining stock returns. In the upcoming section of this paper, a multifactor model incorporating both of those factors and the additional market return (RM) is developed. The general equation representing the return of a stock portfolio which uses these factors as explaining determinants looks as follows:

$$R_P = \alpha_P + c_P \cdot CDS\ Premium + v_P \cdot Volatility\ Premium + \beta_P \cdot Market\ Premium$$

Substituting RMU for the CDS premium factor, VMC for the volatility premium factor, and RM for the market return yields the following regression equation:

$$R_P = \alpha_P + c_P \cdot RMU + v_P \cdot VMC + \beta_P \cdot RM$$

As previously, the equal weighted average returns of the five quintile portfolios sorted by their market to book ratio (MB) are used as independent variables against which the three factors are regressed. The portfolio formation methodology is the same as above. The portfolios are created by sorting the universe into five quintiles according to the individual stock's MB ratio. The portfolios are then regenerated every trading day. The results of regressing the equal weighted average returns of these quintile portfolios against the three factors are shown below:

**Table 11: MB Quintiles regressed against RMU, VMC, and RM**

| | MB Quin~1 | 2 | 3 | 4 | MB Quin~5 |
|---|---|---|---|---|---|
| rmu | 0.493*** | -0.0395 | -0.0836* | -0.161*** | -0.228*** |
| | (10.09) | (-1.00) | (-2.52) | (-5.85) | (-7.64) |
| vmc | 0.111*** | 0.00652 | -0.0523*** | -0.0415*** | -0.0243 |
| | (4.55) | (0.36) | (-3.78) | (-3.55) | (-1.71) |
| rm | 1.023*** | 1.098*** | 0.986*** | 1.012*** | 0.899*** |
| | (38.05) | (39.79) | (41.41) | (58.64) | (39.00) |
| _cons | -0.00000561 | -0.000162 | -0.000165 | 0.000154 | 0.000202 |
| | (-0.02) | (-0.62) | (-0.75) | (0.85) | (0.90) |
| N | 435 | 435 | 435 | 435 | 435 |
| R-sq | 0.972 | 0.958 | 0.950 | 0.962 | 0.940 |

t statistics in parentheses
* p<0.05, ** p<0.01, *** p<0.001

As the above results table indicates, the factor loadings on RMU are positive for quintile one and turn negative for all higher quintile portfolios. This essentially reflects the same results already obtained in the two factor model where the portfolio returns were regressed against RMU and RM only, without VMC. In this three factor model the same characteristics are visible, although the absolute size of the coefficients is smaller than in the two factor model. For example, the factor loading on RMU in quintile one is 0.493 in the three factor model, where as it is 0.681 in the two factor model. For the RMU factor loadings in the higher quintile portfolios, the difference between the two and three factor models is smaller. Even though the absolute size of the coefficients is reduced in the three factor model, almost all factors are statistically highly significant with confidence levels of 99.9% for the quintile one, four, and five portfolios. The only portfolio with a statistically insignificant factor loading on RMU is quintile two. This is however due to the development of positive factor loadings towards negative factor loadings, which cross zero somewhere within the quintile two portfolio and which consequently is very close to zero resulting in a low t-statistic.

The factor loadings on VMC in the three factor regression behave very similar to the factor loadings already obtained in the two factor model. The VMC factor loadings are positive for the lower quintile portfolios and then turn negative for the higher quintile portfolios. This effect is statistically highly significant with some factor loadings having t-statistics beyond 3. The factor loadings for quintiles two and five are however not significant. For the factor loading in quintile two, the explanation for this insignificance seems obvious: Since factor loadings are turning from positive to negative, they have to cross zero somewhere, which happens in the quintile two portfolio. The insignificance of portfolio five is however more intriguing since there seems no ad-hoc explanation for this behavior. The factor loadings on RM, the equal weighted average return of all stocks in this paper’s universe are consistently positive and consistently highly significant with all t-statistics beyond 38.

The three factor model seems to be a highly effective method to explain stock returns since there are no statistically significant intercepts left. All t-statistics of the regression alphas are below 1.0. The abundance of statistically significant alphas is often referred to as the litmus test for an asset pricing model in the world of finance. If this test was to be used on the just developed three factor model, it would pass due to its ability to reduce all intercepts or alphas to statistical insignificance. Also the explanatory power of this three factor model is very high with $R^2$ values ranging from 0.94 to 0.972. Compared to other multifactor models, such as Fama and French’s famous three factor model (1993), which results in $R^2$ values between 0.93 and 0.94, the three factor model proposed in this paper seems to be doing quite a good job to explain stock returns. A regression of the MB quintiles against the original Fama-French factors HML and SMB (French K. , 2010) as well as RM is shown below.

**Table 12: MB Quintiles regressed against HML, SMB, and RM**

| | MB Quin~1 | 2 | 3 | 4 | MB Quin~5 |
|---|---|---|---|---|---|
| hml | 0.957*** | -0.0422 | -0.196*** | -0.328*** | -0.410*** |
| | (10.72) | (-0.99) | (-3.71) | (-7.68) | (-11.19) |
| smb | -0.0504 | -0.0808 | 0.0988* | 0.0616 | -0.0300 |
| | (-0.67) | (-1.64) | (2.13) | (1.58) | (-0.94) |
| rm | 1.314*** | 1.086*** | 0.899*** | 0.916*** | 0.795*** |
| | (25.69) | (50.03) | (29.10) | (38.66) | (43.15) |
| _cons | -0.000119 | -0.000201 | -0.0000431 | 0.000125 | 0.000248 |
| | (-0.22) | (-0.67) | (-0.15) | (0.49) | (0.95) |
| N | 367 | 367 | 367 | 367 | 367 |
| R-sq | 0.944 | 0.959 | 0.938 | 0.950 | 0.929 |

t statistics in parentheses
* p<0.05, ** p<0.01, *** p<0.001

Due to the fact that the Fama-French factors are not available for the last recent months, the sample size is reduced to 367 trading days. As shown above, the factor loading on HML is highly significant and shows a similar behavior as the RMU factor: It starts off with a positive coefficient which then turns negative for all quintile portfolios beyond one. The factor loadings on SMB are almost all statistically non significant, which seems logical since there are only large-cap stocks included in this paper's sample.

Although the HML and the RMU factors are highly correlated with a correlation coefficient of 0.80, they still seem to measure different things. A regression including both has statistically significant positive factor loadings on both factors as table 24 in the appendix shows. The only possible explanation for such a high significance and high correlation other than a true causal difference would be a similar trend in both time series which is picked up by the OLS estimation. However, as a Dickey-Fuller test reveals (see table 25 in the appendix), RMU and HML are stationary and hence it is evident that even though both factors are highly correlated, they still measure different effects at least to a certain extent.

Apart from the original task to develop a working multifactor model based on derivative implied factors, the three factor model based on RMU, VMC, and RM also generates an interesting insight on the often discussed value effect of stock returns which states that stocks with lower MB ratios outperform stocks with higher MB ratios.

When looking at the factor loadings on RMU and VMC, it becomes evident that the findings of the isolated regressions still hold for the complete factor model. Stocks with low MB ratios tend to have positive factor loadings on RMU as well as VMC, whereas stocks with higher MB ratios lean towards negative factor loadings. The chart below shows the development of the factor loadings on RMU and VMC derived from a decile regression which also included RM. The full table of results of this regression can be seen in table 11 in the appendix.

**Picture 7: RMU and VMC Factor Loadings of Decile Regression**

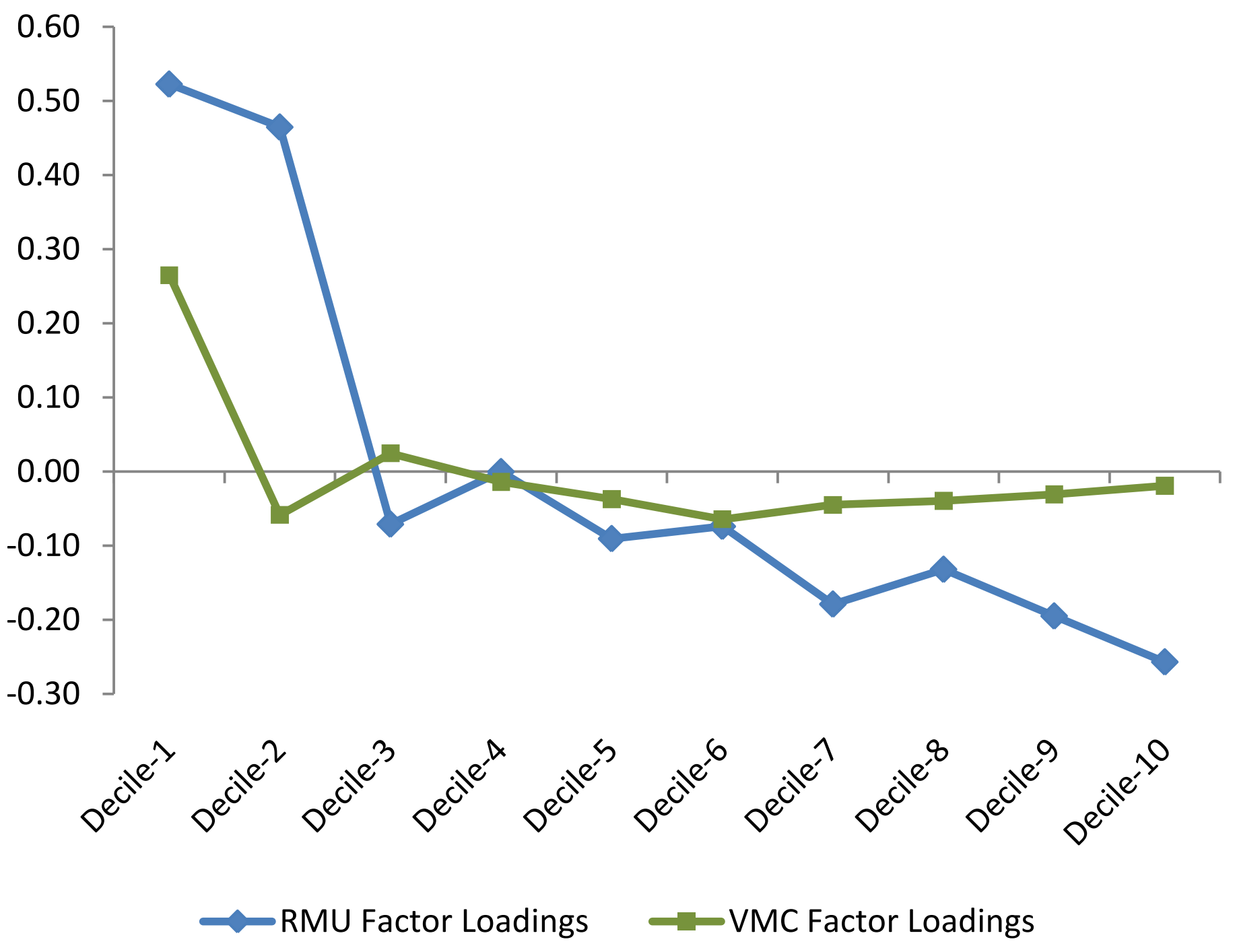

As shown, the factor loadings on both factors are positive for the first decile and turn negative for most of the higher decile portfolios. Also the absolute size of the factor loadings rapidly deteriorates. Especially the VMC factor loadings seem to asymptotically approach zero in the higher decile portfolios, where they are not even statistically different from zero anymore. This implies that the implied volatility premium, which during this observation period was negative, leads to high negative effects on stocks with low MB ratios but only to small positive effects on stocks with high MB ratios. It seems that the implied volatility is a more important topic for distressed securities than it is for securities with higher MB valuations. A similar logic holds true for the RMU factor. Stocks within the lower MB decile portfolios tend to have higher credit spreads as a result of a higher risk of default. Shareholders are compensated for bearing this risk of default by higher stock returns. As soon as the company's credit risk and MB ratio improve, the positive premium earned for bearing credit risk first becomes zero and later even turns negative. The rate of change also massively deteriorates for higher decile portfolios. This means that credit risk is a huge topic for stocks in the lowest MB deciles, but only a minor topic for all other deciles. These effects lead to the outperformance of low MB stocks compared to high MB stocks, commonly known as the value effect.

To conclude: The development of the RMU and VMC factor loadings allows decomposing the value effect into a credit risk premium and a volatility risk premium. Stocks with lower MB ratios have positive factor loadings on RMU and VMC. This means that they are earning a positive return premium for bearing credit risk and a negative return premium for bearing volatility risk. This negative volatility risk premium is however subject to a sample selection bias, which includes a time horizon which is too short to draw any significant conclusions. The findings and implications of the factor loadings on the credit risk premium are more robust since they were essentially the same during the five year observation period.

# 5 Summary

## 5.1 Discussion of Results

The determination of the factor premia has shown that stocks with higher CDS spreads tend to earn a premium over the stocks with lower CDS spreads. Considering classic capital market theories this also seems logical since investors who bear higher risks have to be compensated with higher returns for bearing that risk. During the observation period, the stocks whose CDS spreads were above the 75$^{th}$ percentile outperformed the stocks whose CDS spreads were below the 25$^{th}$ percentile by 5.89% on an annual level. This effect is statistically significant with an extrapolated t-statistic of 9.2 on an annual level.

For the implied volatility the contrary is true. Stocks whose implied volatility is beyond the 95$^{th}$ percentile underperformed stocks whose implied volatility is below the 5$^{th}$ percentile by roughly 79% per year during the observed period. The validity of this finding is however very limited since the observation period for the implied volatilities in this paper is rather short and falls right within the middle of the global financial crisis, during which investors tried to dispose their risky assets, therefore plummeting share prices of stocks with higher implied volatilities. When splitting the observation period into two parts, it becomes evident that during the first 285 observation days the relationship between return and implied volatility was negative but that it was positive for the following 151 observations. This is evidence for a structural break of the risk-return relationship within this time-series which further indicates the necessity of using a larger observation period to confirm the validity of the results obtained in this paper. If this paper's results were valid, they would violate the classic assumption that a higher expected risk should be rewarded by higher expected returns.

The results obtained through regressing the equal weighted average stock returns of various portfolios, which have been formed through sorting the stocks by their MB ratio, have shown that the CDS premium factor, called RMU, is helpful in explaining stock returns. However, the explanatory power and the size of the factor loadings on RMU tend to decrease for portfolios with higher MB ratios. When the MB ratio is seen as an indicator for distress, this leads to an interesting finding. The rapid deterioration of the $R^2$ and the factor loadings on RMU reflects the fact that credit risk is only important to shareholders when it is very high. As soon as the credit risk gets lower and the MB ratios correspondingly increase, credit risk becomes almost irrelevant to investors. It also becomes evident that the so-called value effect is not an anomaly in market behavior but actually a logical consequence of credit risk. Stock holders bearing a risk of losing their whole investment in the case of a credit event are being compensated for that risk. This risk premium does, at least partially, explain the value effect of stock returns. It is further very interesting to see that this effect is almost entirely created by the lowest quintile or decile portfolios, which account for a majority of the difference in returns. This is reflected in *high positive* factor loadings for low MB quintile stocks but only *slightly negative* factor loadings for high MB quintile stocks.

The regression analyses on the VMC factor, which controls for the implied volatilities, have shown that stocks with low MB ratios tend to have more implied volatility compared to the market average. Since above average implied volatility was rewarded with negative return premiums during the observed time period, the stocks in the low MB quintile portfolios earn negative return premiums compared to stocks with higher

MB ratios which have less implied volatility. The factor loadings on VMC are especially high for the lower MB decile or quintile portfolios and turn negative for the higher portfolios. The absolute size of the factor loadings is also much higher for stocks with lower MB ratios. This is evidence that the implied volatility is a more important determinant of stock returns for stocks with high implied volatility than for stocks with low volatility. This means that stock returns are largely affected by high implied volatilities in the low MB sector, but that low volatilities on the other hand have only minor effects on stock returns.

In the full three factor regression model, which includes both derivative implied factors (RMU and VMC) as well as the market return (RM), it becomes evident that the CDS spread as well as the implied volatility are indeed helpful in explaining stock returns. Most of the factor loadings on the explaining factors are statistically highly significant, except for those quintiles in which the factor loadings turn from positive into negative territory. The positive factor loadings on RMU and VMC in the lower MB portfolios show that low MB ratios are correlated with higher credit spreads and higher implied volatilities. For stocks with higher MB ratios these factor loadings decrease and become even negative for stocks with the highest MB ratios. This indicates that low MB stocks earn a premium for higher credit risk and a discount for higher implied volatility. As the MB ratio increases, the credit spreads and implied volatility decrease and so do the factor loadings on the RMU and VMC factors. The interesting fact is not that the factor loadings turn from positive to negative, but that the absolute value of the factor loadings for lower MB stocks is much higher than the absolute value for higher MB stocks. This shows that credit risk and volatility are more important determinants of a stock's return when the credit risk and implied volatility are high, than for stocks with lower risks. It seems that as soon as company's credit risk and its stock's implied volatility have fallen below a certain threshold, these two variables lose a large portion of their explanatory power. Especially for credit risk this finding makes sense since the risk of default is only relevant for a shareholder if there is a high probability that a company defaults in which case the shareholder would lose his investment. If the credit risk is already on a very low level, shareholders focus their attention on topics other than credit risk.

Also the explanatory power of this three factor model is very high with $R^2$ values between 0.94 and 0.97 in the quintile regression analysis. This is even slightly higher than the $R^2$ values achieved in the Fama-French three factor analysis. Also the fact that all intercepts, also often referred to as alphas, are statistically insignificant. This insignificance in combination with the high explanatory power which explains almost all variation in stock returns underlines the usefulness of this multifactor model based on derivative data. It seems that the credit default swap spreads as well as the implied volatility are useful factors in explaining stock returns.

In addition to the finding that the three factor model is useful for explaining stock returns, this paper also makes interesting findings regarding the value effect of stock returns by observing the development of the factor loadings in relationship to the MB ratio. The different factor loadings on RMU and VMC for the specific MB portfolios show that the value effect could, at least partially, be attributed to a credit risk premium and possibly an implied volatility premium. Stocks with lower MB ratios bear excess credit risk which is being compensated by a positive premium on stock returns. This positive premium for low MB stocks is larger than the negative premium earned by high MB stocks. Hence, the value effect is not an anomaly of capital markets but adequately expresses a positive return premium for credit risk which is inflicted upon an investor of low MB stocks. It is likely that a similar conclusion could be drawn for the implied volatility, but due to the already discussed sample bias which includes a too short time horizon, a larger sample is needed for further analysis.

The findings of this paper also have some practical implications. The possibility to explain stock returns through credit market derivatives could potentially allow for cross-hedging the exposures of these different asset classes. Especially a long exposure in stocks with high credit risk could be partially hedged through a replication of the RMU portfolio or a direct investment in CDS. The econometric regression model shown above could be modified in order to calculate the exact hedge ratios needed for an optimal hedge. The presented relationship between the stock and the credit market would also allow improving risk assessment systems through incorporating possible correlation and cluster risks which affect stock as well as credit markets.

Another advantage of the proposed three factor model is that the credit risk and the implied volatility are very sluggish and can hence be predicted very accurately over a short-term horizon. The possibility of predicting these two factors could theoretically allow creating profitable trading strategies based on the RMU and VMC portfolio.

## 5.2 Ideas for Further Research

Especially the small observation period for the implied volatility factor is a problem that should be avoided in further studies. In order to make accurate and reliable statements about the relationship between stock returns and the implied volatility, a larger sample size is needed. If possible, the time horizon should include several full economic cycles and thus be at least 10 years long. The only limitation to the time horizon is the availability of data on credit default swaps and equity options, which are used to calculate the implied volatility.

Additionally, further studies should not only try to use a longer time period but also to include a larger stock universe. Using a larger universe would allow to include more risk factors such as the size of a company in the regression analysis. Since all S&P 100 stocks used as universe in this paper are large-cap stocks, size effects are virtually not analyzable by using the universe of in this paper. By including additional stocks, it would be very interesting to analyze the factor loadings on RMU and VMC of portfolios which have been created with size or market cap as sorting criteria instead of the market to book ratio. A disadvantage however of using a larger universe is the fact that the availability and liquidity of derivatives is much smaller for small-cap stocks than it is for large-cap stocks.

Another potential idea for research would be to decompose the Fama-French HML factor, also known as value factor, into credit risk and implied volatility risk. If the conclusion of this paper is correct, it should be possible to explain a large portion of this HML factor through credit risk and implied volatility. Also a comparison of the three factor model developed in this paper and its replication using cash market bond spreads instead of CDS spreads would be interesting to conduct. In terms of the econometric methods used, a cross-sectional test of the results presented in this paper analogue to the procedure used by Black, Jensen, and Scholes (1972) could reveal further insights.

## 5.3 Conclusion

This paper has shown that derivative markets can be used to build multifactor asset pricing models. The CDS spread and the implied volatility of a stock are helpful determinants in explaining the return behavior of stocks. Especially stocks with high credit risk and high implied volatility are largely affected by the premiums earned on these two factors. Stocks with low credit risk and low implied volatility are almost unaffected by these two factors. Since credit risk and volatility are correlated with the market to book ratio of a stock, it can be concluded that the often discussed value effect is not an effect of a market anomaly or irrational investing behavior: The value effect is actually a consequence of the credit risk and the implied volatility of a company. If a company has a low MB ratio, its credit risk and its implied volatility are very high and the shareholder is in imminent danger of losing his investment. That risk is being rewarded with positive stock return premiums. As soon as the MB ratio and with it the credit risk and the implied volatility improve, the factor loadings on the risk premia largely disappear. As soon as a certain threshold is passed, investors stop worrying about credit risk and implied volatility and probably focus their attention on different aspects, such as corporate strategy and the overall economic development.

It can therefore be concluded that the model developed in this paper is a useful expansion of previously developed multifactor models. The high statistical significance of the factor loadings and the large overall explanatory power of the proposed three factor model based on market return, credit default swap spreads, and implied volatility make it clear that derivative markets are very useful in developing asset pricing models and estimating risk premia.

## Bibliography

Black, F., & Scholes, M. (1973). The Pricing of Options and Corporate Liabilities. *The Journal of Political Economy* , 637-654.

Black, F., Jensen, M. C., & Scholes, M. (1972). The Capital Asset Pricing Model: Some Empirical Tests. *Studies in the Theory of Capital Markets* , 79–121.

Campbell, J. Y., & Hentschel, L. (1992). No News is Good News - An Asymmetric Model of Changing Volatility in Stock Returns. *Journal of Financial Economics* , 281-318.

Campbell, J. Y., & Shiller, R. J. (1988). Stock Prices, Earnings, and Expected Dividends. *The Journal of Finance* , 661-676.

Carhart, M. M. (1997). On Persistence in Mutual Fund Performance. *The Journal of Finance* , 57-82.

Carr, P., & Wu, L. (2009). Stock Options and Credit Default Swaps: A Joint Framework for Valuation and Estimation. *Journal of Financial Econometrics* , 1-41.

Chen, N.-F., Roll, R., & Ross, S. A. (1986). Economic Forces and the Stock Market. *The Journal of Business* , 383-403.

Cox, J. C., Ingersoll, J. E., & Ross, S. A. (1985). An Intertemporal General Equilibrium Model of Asset Prices. *Econometrica* , 363-384.

Daniels, K. N., & Jensen, M. S. (2005). The Effect of Credit Ratings on Credit Default Swap Spreads and Credit Spreads. *Journal of Fixed Income* , 16-33.

DeBondt, W. F., & Thaler, R. (1985). Does the Stock Market Overreact? *The Journal of Finance* , 793-805.

DeLong, J. B., Shleifer, A., Summers, L. H., & Waldman, R. J. (1990). Positive Feedback Investment Strategies and Destabilizing Rational Speculation. *The Journal of Finance* , 379-395.

Derman, E., & Kani, I. (1994). Riding on a Smile. *Risk* , 32-39.

Dupire, B. (1994). Pricing with a Smile. *Risk* , 18-20.

European Central Bank. (2009). *Credit Default Swaps and Counterparty Risk.* Frankfurt: European Central Bank.

Fabozzi, F. J. (2005). *The Handbook of Fixed Income Securities.* New York: McGraw-Hill.

Fama, E. F., & French, K. R. (1989). Business Conditions and Expected Returns on Stocks and Bonds. *Journal of Financial Economics* , 23-49.

Fama, E. F., & French, K. R. (1993). Common Risk Factors in the Returns of Stocks and Bonds. *Journal of Financial Economics* , 3-56.

Fama, E. F., & French, K. R. (1992). The Cross-Section of Expected Stock Returns. *The Journal of Finance* , 427-465.

Fama, E. F., & French, K. R. (1998). Value Versus Growth: The International Evidence. *The Journal of Finance* , 1975-1999.

French, K. (2010, January 1). *Fama French Factors*. Retrieved January 1, 2010, from http://mba.tuck.dartmouth.edu/pages/faculty/ken.french/data_library.html

French, K. R., Schwert, G. W., & Stambaugh, R. F. (1986). Expected Stock Returns and Volatility. *Journal of Financial Economics* , 3-29.

Glosten, L. R., Jagannathan, R., & Runkle, D. E. (1993). On the Relation between the Expected Value and the Volatility of the Nominal Excess Return on Stocks. *The Journal of Finance* , 1779-1801.

Hull, J. C. (2009). *Options, Futures, and Other Derivatives.* New Jersey: Pearson Prentice Hall.

International Swaps and Derivatives Association. (2010, January 12). ISDA CDS Master Agreement. New York, NY, United States of America: ISDA. Retrieved from ISDA CDS Master Agreement: http://www.isda.org/cgi-bin/_isdadocsdownload/download.asp?DownloadID=84

Jegadeesh, N., & Titman, S. (1993). Returns to Buying Winners and Selling Losers: Implications for Stock Market Efficiency. *The Journal of Finance* , 65-91.

Jones, C. M., & Lamont, O. A. (2002). Short-Sale Constraints and Stock Returns. *Journal of Financial Economics* , 207-239.

Kawaller, I. G., Koch, P. D., & Koch, T. W. (1987). The Temporal Price Relationship between S&P 500 Futures and the S&P 500 Index. *The Journal of Finance* , 1309-1329.

Lamont, O. (1998). Earnings and Expected Returns. *The Journal of Finance* , 1563-1587.

Lintner, J. (1965). The Valuation of Risk Assets and the Selection of Risky Investments in Stock. *The Review of Economics and Statistics* , 13-37.

Longstaff, F. A., & Schwartz, E. S. (2001). Valuing American Options by Simulation: A Simple Least-Squares Approach. *The Review of Financial Studies* , 113-147.

Markowitz, H. (1952). Portfolio Selection. *The Journal of Finance* , 77-91.

Merton, R. C. (1973). An Intertemporal Capital Asset Pricing Model. *Econometrica* , 867-887.

Merton, R. C. (1973). Theory of Rational Option Pricing. *The Bell Journal of Economics and Management Science* , 141-183.

Norden, L., & Weber, M. (2009). The Co-movement of Credit Default Swap, Bond, and Stock Markets: An Empirical Analysis. *European Financial Management* , 529-562.

Ross, S. A. (1976). The Arbitrage Theory of Capital Asset Pricing. *Journal of Economic Theory* , 341-360.

Rubinstein, M. (1994). Implied Binomial Trees. *The Journal of Finance* , 771-818.

Sharpe, W. F. (1964). Capital Asset Prices: A Theory of Market Equillibrium Under Conditions of Risk. *The Journal of Finance* , 425-442.

Sharpe, W. F. (1966). Mutual Fund Performance. *Journal of Business* , 119-138.

Sortino, F., & Van der Meer, R. (1991). Downside Risk. *The Journal of Portfolio Management* , 27-31.

US Securities and Exchange Commission. (2008, September 23th). *Testimony Concerning Turmoil in U.S. Credit Markets: Recent Actions Regarding Government Sponsored Entities, Investment Banks and Other Financial Institutions*. Retrieved January 4th, 2010, from US Securities and Exchange Commission: http://www.sec.gov/news/testimony/2008/ts092308cc.htm

Zhang, B. Y., Zhou, H., & Zhu, H. (2009). Explaining Credit Default Swap Spreads with the Equity Volatility and Jump Risks of Individual Firms. *The Review of Financial Studies* , 5099-5131.

# Appendix

| Symbol | Variable | Definition or Source |
|---|---|---|
| | **Basic Series** | |
| I | Inflation | Log relative of U.S. Consumer Price Index |
| TB | Treasury-bill rate | End-of-period return on 1-month bills |
| LGB | Long-term government bonds | Return on long-term government bonds (1958–78: Ibbotson and Sinquefield [1982]; 1979–83: CRSP) |
| IP | Industrial production | Industrial production during month (*Survey of Current Business*) |
| Baa | Low-grade bonds | Return on bonds rated Baa and under (1953–77: Ibbotson [1979], constructed for 1978–83) |
| EWNY | Equally weighted equities | Return on equally weighted portfolio of NYSE-listed stocks (CRSP) |
| VWNY | Value-weighted equities | Return on a value-weighted portfolio of NYSE-listed stocks (CRSP) |
| CG | Consumption | Growth rate in real per capita consumption (Hansen and Singleton [1982]; *Survey of Current Business*) |
| OG | Oil prices | Log relative of Producer Price Index/Crude Petroleum series (Bureau of Labor Statistics) |
| | **Derived Series** | |
| $MP(t)$ | Monthly growth, industrial production | $\log_e[IP(t)/IP(t-1)]$ |
| $YP(t)$ | Annual growth, industrial production | $\log_e[IP(t)/IP(t-12)]$ |
| $E[I(t)]$ | Expected inflation | Fama and Gibbons (1984) |
| $UI(t)$ | Unexpected inflation | $I(t) - E[I(t)\|t-1]$ |
| $RHO(t)$ | Real interest (ex post) | $TB(t-1) - I(t)$ |
| $DEI(t)$ | Change in expected inflation | $E[I(t+1)\|t] - E[I(t)\|t-1]$ |
| $URP(t)$ | Risk premium | $Baa(t) - LGB(t)$ |
| $UTS(t)$ | Term structure | $LGB(t) - TB(t-1)$ |

**Table 13: Risk Factors (Chen, Roll, & Ross, 1986)**

Regressions of excess stock and bond returns (in percent) on the excess stock-market return, $RM-RF$: July 1963 to December 1991, 342 months.[a]

$$R(t) - RF(t) = a + b[RM(t) - RF(t)] + e(t)$$

Dependent variable: Excess returns on 25 stock portfolios formed on size and book-to-market equity

| Size quintile | Book-to-market equity ($BE/ME$) quintiles | | | | | | | | | |
|---|---|---|---|---|---|---|---|---|---|---|
| | Low | 2 | 3 | 4 | High | Low | 2 | 3 | 4 | High |
| | $b$ | | | | | $t(b)$ | | | | |
| Small | 1.40 | 1.26 | 1.14 | 1.06 | 1.08 | 26.33 | 28.12 | 27.01 | 25.03 | 23.01 |
| 2 | 1.42 | 1.25 | 1.12 | 1.02 | 1.13 | 35.76 | 35.56 | 33.12 | 33.14 | 29.04 |
| 3 | 1.36 | 1.15 | 1.04 | 0.96 | 1.08 | 42.98 | 42.52 | 37.50 | 35.81 | 31.16 |
| 4 | 1.24 | 1.14 | 1.03 | 0.98 | 1.10 | 51.67 | 55.12 | 46.96 | 37.00 | 32.76 |
| Big | 1.03 | 0.99 | 0.89 | 0.84 | 0.89 | 51.92 | 61.51 | 43.03 | 35.96 | 27.75 |
| | $R^2$ | | | | | $s(e)$ | | | | |
| Small | 0.67 | 0.70 | 0.68 | 0.65 | 0.61 | 4.46 | 3.76 | 3.55 | 3.56 | 3.92 |
| 2 | 0.79 | 0.79 | 0.76 | 0.76 | 0.71 | 3.34 | 2.96 | 2.85 | 2.59 | 3.25 |
| 3 | 0.84 | 0.84 | 0.80 | 0.79 | 0.74 | 2.65 | 2.28 | 2.33 | 2.26 | 2.90 |
| 4 | 0.89 | 0.90 | 0.87 | 0.80 | 0.76 | 2.01 | 1.73 | 1.84 | 2.21 | 2.83 |
| Big | 0.89 | 0.92 | 0.84 | 0.79 | 0.69 | 1.66 | 1.35 | 1.73 | 1.95 | 2.69 |

**Table 14: CAPM Regression (Fama & French, 1993, p. 20)**

Regressions of excess stock and bond returns (in percent) on the excess market return ($RM-RF$) and the mimicking returns for the size ($SMB$) and book-to-market equity ($HML$) factors: July 1963 to December 1991, 342 months.[a]

$$R(t) - RF(t) = a + b[RM(t) - RF(t)] + sSMB(t) + hHML(t) + e(t)$$

Dependent variable: Excess returns on 25 stock portfolios formed on size and book-to-market equity

Book-to-market equity ($BE/ME$) quintiles

| Size quintile | Low | 2 | 3 | 4 | High | Low | 2 | 3 | 4 | High |
|---|---|---|---|---|---|---|---|---|---|---|
| | | | $b$ | | | | | $t(b)$ | | |
| Small | 1.04 | 1.02 | 0.95 | 0.91 | 0.96 | 39.37 | 51.80 | 60.44 | 59.73 | 57.89 |
| 2 | 1.11 | 1.06 | 1.00 | 0.97 | 1.09 | 52.49 | 61.18 | 55.88 | 61.54 | 65.52 |
| 3 | 1.12 | 1.02 | 0.98 | 0.97 | 1.09 | 56.88 | 53.17 | 50.78 | 54.38 | 52.52 |
| 4 | 1.07 | 1.08 | 1.04 | 1.05 | 1.18 | 53.94 | 53.51 | 51.21 | 47.09 | 46.10 |
| Big | 0.96 | 1.02 | 0.98 | 0.99 | 1.06 | 60.93 | 56.76 | 46.57 | 53.87 | 38.61 |
| | | | $s$ | | | | | $t(s)$ | | |
| Small | 1.46 | 1.26 | 1.19 | 1.17 | 1.23 | 37.92 | 44.11 | 52.03 | 52.85 | 50.97 |
| 2 | 1.00 | 0.98 | 0.88 | 0.73 | 0.89 | 32.73 | 38.79 | 34.03 | 31.66 | 36.78 |
| 3 | 0.76 | 0.65 | 0.60 | 0.48 | 0.66 | 26.40 | 23.39 | 21.23 | 18.62 | 21.91 |
| 4 | 0.37 | 0.33 | 0.29 | 0.24 | 0.41 | 12.73 | 11.11 | 9.81 | 7.38 | 11.01 |
| Big | − 0.17 | − 0.12 | − 0.23 | − 0.17 | − 0.05 | − 7.18 | − 4.51 | − 7.58 | − 6.27 | − 1.18 |
| | | | $h$ | | | | | $t(h)$ | | |
| Small | − 0.29 | 0.08 | 0.26 | 0.40 | 0.62 | − 6.47 | 2.35 | 9.66 | 15.53 | 22.24 |
| 2 | − 0.52 | 0.01 | 0.26 | 0.46 | 0.70 | − 14.57 | 0.41 | 8.56 | 17.24 | 24.80 |
| 3 | − 0.38 | − 0.00 | 0.32 | 0.51 | 0.68 | − 11.26 | − 0.05 | 9.75 | 16.88 | 19.39 |
| 4 | − 0.42 | 0.04 | 0.30 | 0.56 | 0.74 | − 12.51 | 1.04 | 8.83 | 14.84 | 17.09 |
| Big | − 0.46 | 0.00 | 0.21 | 0.57 | 0.76 | − 17.03 | 0.09 | 5.80 | 18.34 | 16.24 |
| | | | $R^2$ | | | | | $s(e)$ | | |
| Small | 0.94 | 0.96 | 0.97 | 0.97 | 0.96 | 1.94 | 1.44 | 1.16 | 1.12 | 1.22 |
| 2 | 0.95 | 0.96 | 0.95 | 0.95 | 0.96 | 1.55 | 1.27 | 1.31 | 1.16 | 1.23 |
| 3 | 0.95 | 0.94 | 0.93 | 0.93 | 0.93 | 1.45 | 1.41 | 1.43 | 1.32 | 1.52 |
| 4 | 0.94 | 0.93 | 0.91 | 0.89 | 0.89 | 1.46 | 1.48 | 1.49 | 1.63 | 1.88 |
| Big | 0.94 | 0.92 | 0.88 | 0.90 | 0.83 | 1.16 | 1.32 | 1.55 | 1.36 | 2.02 |

**Table 15: Fama-French - RMRF, SMB, HML (Fama & French, 1993, pp. 23-24)**

Regressions of excess stock and bond returns (in percent) on the bond-market returns, *TERM* and *DEF*: July 1963 to December 1991, 342 months.[a]

$$R(t) - RF(t) = a + mTERM(t) + dDEF(t) + e(t)$$

Dependent variable: Excess returns on 25 stock portfolios formed on size and book-to-market equity

| | Book-to-market equity (*BE/ME*) quintiles | | | | | | | | | |
|---|---|---|---|---|---|---|---|---|---|---|
| Size quintile | Low | 2 | 3 | 4 | High | Low | 2 | 3 | 4 | High |
| | $m$ | | | | | $t(m)$ | | | | |
| Small | 0.93 | 0.90 | 0.89 | 0.86 | 0.89 | 5.02 | 5.50 | 5.95 | 6.08 | 6.01 |
| 2 | 0.99 | 0.96 | 0.99 | 1.01 | 0.98 | 5.71 | 6.32 | 7.29 | 8.34 | 6.92 |
| 3 | 0.99 | 0.94 | 0.94 | 0.95 | 0.99 | 6.25 | 7.10 | 7.80 | 8.50 | 7.60 |
| 4 | 0.92 | 0.95 | 0.97 | 1.05 | 1.03 | 6.58 | 7.57 | 8.53 | 9.64 | 7.83 |
| Big | 0.82 | 0.82 | 0.80 | 0.80 | 0.77 | 7.14 | 7.60 | 8.09 | 8.26 | 6.84 |
| | $d$ | | | | | $t(d)$ | | | | |
| Small | 1.39 | 1.31 | 1.33 | 1.45 | 1.52 | 3.96 | 4.27 | 4.73 | 5.45 | 5.45 |
| 2 | 1.26 | 1.28 | 1.35 | 1.38 | 1.41 | 3.84 | 4.47 | 5.28 | 6.05 | 5.29 |
| 3 | 1.21 | 1.19 | 1.25 | 1.24 | 1.21 | 4.05 | 4.74 | 5.49 | 5.89 | 4.88 |
| 4 | 0.96 | 1.01 | 1.13 | 1.21 | 1.22 | 3.65 | 4.28 | 5.25 | 5.89 | 4.92 |
| Big | 0.78 | 0.73 | 0.78 | 0.83 | 0.89 | 3.59 | 3.60 | 4.18 | 4.56 | 4.15 |
| | $R^2$ | | | | | $s(e)$ | | | | |
| Small | 0.06 | 0.08 | 0.09 | 0.10 | 0.10 | 7.50 | 6.57 | 6.00 | 5.68 | 5.95 |
| 2 | 0.08 | 0.10 | 0.13 | 0.17 | 0.12 | 6.97 | 6.09 | 5.45 | 4.87 | 5.69 |
| 3 | 0.10 | 0.12 | 0.15 | 0.17 | 0.14 | 6.38 | 5.35 | 4.86 | 4.48 | 5.28 |
| 4 | 0.11 | 0.14 | 0.17 | 0.21 | 0.15 | 5.63 | 5.04 | 4.57 | 4.39 | 5.31 |
| Big | 0.13 | 0.15 | 0.16 | 0.17 | 0.12 | 4.61 | 4.33 | 4.00 | 3.89 | 4.55 |

**Table 16: Fama-French - TERM, DEF (Fama & French, 1993, p. 17)**

Regressions of excess stock returns on 25 stock portfolios formed on size and book-to-market equity (in percent) on the stock-market returns, *RM–RF*, *SMB*, and *HML*, and the bond-market returns, *TERM* and *DEF*: July 1963 to December 1991, 342 months.[a]

$$R(t) - RF(t) = a + b[RM(t) - RF(t)] + sSMB(t) + hHML(t) + mTERM(t) + dDEF(t) + e(t)$$

| Size quintile | Book-to-market equity (BE/ME) quintiles | | | | | | | | | |
|---|---|---|---|---|---|---|---|---|---|---|
| | Low | 2 | 3 | 4 | High | Low | 2 | 3 | 4 | High |
| | | | *b* | | | | | *t(b)* | | |
| Small | 1.06 | 1.04 | 0.96 | 0.92 | 0.98 | 35.97 | 47.65 | 54.48 | 54.51 | 53.15 |
| 2 | 1.12 | 1.06 | 0.98 | 0.94 | 1.10 | 47.19 | 54.95 | 49.01 | 54.19 | 59.00 |
| 3 | 1.13 | 1.01 | 0.97 | 0.95 | 1.08 | 50.93 | 46.95 | 44.57 | 47.59 | 46.92 |
| 4 | 1.07 | 1.07 | 1.01 | 1.00 | 1.17 | 48.18 | 47.55 | 44.83 | 41.02 | 41.02 |
| Big | 0.96 | 1.02 | 0.98 | 1.00 | 1.10 | 53.87 | 51.01 | 41.35 | 48.29 | 35.96 |
| | | | *s* | | | | | *t(s)* | | |
| Small | 1.45 | 1.26 | 1.20 | 1.15 | 1.21 | 37.02 | 43.42 | 50.89 | 51.36 | 49.55 |
| 2 | 1.01 | 0.98 | 0.89 | 0.74 | 0.89 | 32.06 | 38.10 | 33.68 | 32.12 | 35.79 |
| 3 | 0.76 | 0.66 | 0.60 | 0.49 | 0.68 | 25.82 | 22.97 | 20.83 | 18.54 | 22.32 |
| 4 | 0.38 | 0.34 | 0.30 | 0.26 | 0.42 | 12.71 | 11.36 | 9.99 | 8.05 | 11.07 |
| Big | − 0.17 | − 0.11 | − 0.23 | − 0.17 | − 0.06 | − 7.03 | − 4.07 | − 7.31 | − 6.07 | − 1.44 |
| | | | *h* | | | | | *t(h)* | | |
| Small | − 0.27 | 0.10 | 0.27 | 0.40 | 0.63 | − 5.95 | 2.90 | 9.82 | 15.47 | 22.27 |
| 2 | − 0.51 | 0.02 | 0.25 | 0.44 | 0.71 | − 14.01 | 0.69 | 8.11 | 16.50 | 24.61 |
| 3 | − 0.37 | − 0.00 | 0.31 | 0.50 | 0.69 | − 10.81 | − 0.11 | 9.28 | 16.18 | 19.34 |
| 4 | − 0.42 | 0.04 | 0.29 | 0.53 | 0.75 | − 12.09 | 1.10 | 8.37 | 14.20 | 16.88 |
| Big | − 0.46 | 0.01 | 0.21 | 0.58 | 0.78 | − 16.85 | 0.38 | 5.70 | 18.16 | 16.59 |
| | | | *m* | | | | | *t(m)* | | |
| Small | − 0.10 | − 0.11 | − 0.05 | − 0.04 | − 0.06 | − 1.93 | − 2.70 | − 1.49 | − 1.19 | − 1.87 |
| 2 | − 0.05 | − 0.04 | 0.07 | 0.14 | − 0.05 | − 1.16 | − 1.12 | 1.90 | 4.33 | − 1.48 |
| 3 | − 0.04 | 0.02 | 0.06 | 0.09 | 0.01 | − 0.91 | 0.53 | 1.48 | 2.48 | 0.25 |
| 4 | − 0.02 | 0.00 | 0.08 | 0.18 | − 0.01 | − 0.55 | 0.19 | 1.92 | 3.98 | − 0.19 |
| Big | 0.03 | − 0.04 | − 0.00 | − 0.04 | − 0.16 | 0.82 | − 0.98 | − 0.06 | − 0.98 | − 2.82 |
| | | | *d* | | | | | *t(d)* | | |
| Small | − 0.17 | − 0.19 | − 0.10 | 0.06 | 0.02 | − 1.74 | − 2.70 | − 1.76 | 1.06 | 0.34 |
| 2 | − 0.12 | − 0.11 | 0.04 | 0.15 | − 0.07 | − 1.59 | − 1.83 | 0.61 | 2.64 | − 1.24 |
| 3 | − 0.09 | − 0.01 | 0.07 | 0.10 | − 0.16 | − 1.25 | − 0.17 | 1.00 | 1.51 | − 2.11 |
| 4 | − 0.11 | − 0.10 | 0.04 | 0.13 | − 0.12 | − 1.51 | − 1.44 | 0.59 | 1.64 | − 1.30 |
| Big | 0.06 | − 0.14 | − 0.02 | − 0.07 | − 0.18 | 0.97 | − 2.15 | − 0.25 | − 1.08 | − 1.84 |
| | | | $R^2$ | | | | | *s(e)* | | |
| Small | 0.94 | 0.96 | 0.97 | 0.97 | 0.96 | 1.93 | 1.43 | 1.16 | 1.11 | 1.20 |
| 2 | 0.95 | 0.96 | 0.95 | 0.95 | 0.96 | 1.55 | 1.27 | 1.31 | 1.13 | 1.23 |
| 3 | 0.95 | 0.94 | 0.93 | 0.93 | 0.93 | 1.45 | 1.41 | 1.43 | 1.31 | 1.50 |
| 4 | 0.94 | 0.93 | 0.91 | 0.90 | 0.89 | 1.46 | 1.47 | 1.48 | 1.59 | 1.88 |
| Big | 0.94 | 0.92 | 0.87 | 0.90 | 0.83 | 1.17 | 1.31 | 1.55 | 1.36 | 2.00 |

**Table 17: Fama-French - RMRF, SML, HML, TERM, DEF (Fama & French, 1993, pp. 28-29)**

**Performance Measurement Model Summary Statistics, July 1963 to December 1993**

VWRF is the Center for Research in Security Prices (CRSP) value-weight stock index minus the one-month T-bill return. RMRF is the excess return on Fama and French's (1993) market proxy. SMB and HML are Fama and French's factor-mimicking portfolios for size and book-to-market equity. PR1YR is a factor-mimicking portfolio for one-year return momentum.

| Factor Portfolio | Monthly Excess Return | Std Dev | $t$-stat for Mean = 0 | Cross-Correlations | | | | |
|---|---|---|---|---|---|---|---|---|
| | | | | VWRF | RMRF | SMB | HML | PR1YR |
| VWRF | 0.44 | 4.39 | 1.93 | 1.00 | | | | |
| RMRF | 0.47 | 4.43 | 2.01 | 1.00 | 1.00 | | | |
| SMB | 0.29 | 2.89 | 1.89 | 0.35 | 0.32 | 1.00 | | |
| HML | 0.46 | 2.59 | 3.42 | −0.36 | −0.37 | 0.10 | 1.00 | |
| PR1YR | 0.82 | 3.49 | 4.46 | 0.01 | 0.01 | −0.29 | −0.16 | 1.00 |

**Table 18: Carhart - Summary Statistics (Carhart, 1997, p. 77)**

**Table III**

## Portfolios of Mutual Funds Formed on Lagged 1-Year Return

Mutual funds are sorted on January 1 each year from 1963 to 1993 into decile portfolios based on their previous calendar year's return. The portfolios are equally weighted monthly so the weights are readjusted whenever a fund disappears. Funds with the highest past one-year return comprise decile 1 and funds with the lowest comprise decile 10. Deciles 1 and 10 are further subdivided into thirds on the same measure. VWRF is the excess return on the CRSP value-weight market proxy. RMRF, SMB, and HML are Fama and French's (1993) market proxy and factor-mimicking portfolios for size and book-to-market equity. PR1YR is a factor-mimicking portfolio for one-year return momentum. Alpha is the intercept of the Model. The t-statistics are in parentheses.

| | | | CAPM | | | 4-Factor Model | | | | | |
|---|---|---|---|---|---|---|---|---|---|---|---|
| Portfolio | Monthly Excess Return | Std Dev | Alpha | VWRF | Adj R-sq | Alpha | RMRF | SMB | HML | PR1YR | Adj R-Sq |
| 1A | 0.75% | 5.45% | 0.27% | 1.08 | 0.777 | −0.11% | 0.91 | 0.72 | −0.07 | 0.33 | 0.891 |
| | | | (2.06) | (35.94) | | (−1.11) | (37.67) | (19.95) | (−1.65) | (11.53) | |
| 1B | 0.67% | 4.94% | 0.22% | 1.00 | 0.809 | −0.10% | 0.86 | 0.59 | −0.05 | 0.27 | 0.898 |
| | | | (2.00) | (39.68) | | (−1.08) | (40.66) | (18.47) | (−1.38) | (10.63) | |
| 1C | 0.63% | 4.95% | 0.17% | 1.02 | 0.843 | −0.15% | 0.89 | 0.56 | −0.05 | 0.27 | 0.927 |
| | | | (1.70) | (44.65) | | (−1.92) | (49.76) | (20.86) | (−1.61) | (12.69) | |
| 1 (high) | 0.68% | 5.04% | 0.22% | 1.03 | 0.834 | −0.12% | 0.88 | 0.62 | −0.05 | 0.29 | 0.933 |
| | | | (2.10) | (43.11) | | (−1.60) | (50.54) | (23.67) | (−1.86) | (13.88) | |
| 2 | 0.59% | 4.72% | 0.14% | 1.01 | 0.897 | −0.10% | 0.89 | 0.46 | −0.05 | 0.20 | 0.955 |
| | | | (1.75) | (57.00) | | (−1.78) | (66.47) | (22.95) | (−2.25) | (12.43) | |
| 3 | 0.43% | 4.56% | −0.01% | 0.99 | 0.931 | −0.18% | 0.90 | 0.34 | −0.07 | 0.16 | 0.963 |
| | | | (−0.08) | (70.96) | | (−3.65) | (76.80) | (18.99) | (−3.69) | (11.52) | |
| 4 | 0.45% | 4.41% | 0.02% | 0.97 | 0.952 | −0.12% | 0.90 | 0.27 | −0.05 | 0.11 | 0.971 |
| | | | (0.33) | (85.70) | | (−2.81) | (90.03) | (18.18) | (−3.12) | (9.40) | |
| 5 | 0.38% | 4.35% | −0.05% | 0.96 | 0.960 | −0.14% | 0.90 | 0.22 | −0.05 | 0.07 | 0.970 |
| | | | (−1.10) | (93.93) | | (−3.31) | (89.65) | (14.42) | (−3.27) | (6.18) | |
| 6 | 0.40% | 4.36% | −0.02% | 0.96 | 0.958 | −0.12% | 0.90 | 0.22 | −0.04 | 0.08 | 0.968 |
| | | | (−0.46) | (91.94) | | (−2.82) | (86.16) | (14.02) | (−2.37) | (6.01) | |
| 7 | 0.36% | 4.30% | −0.06% | 0.95 | 0.959 | −0.14% | 0.90 | 0.21 | −0.03 | 0.04 | 0.967 |
| | | | (−1.39) | (92.90) | | (−3.09) | (85.73) | (13.17) | (−1.62) | (2.89) | |
| 8 | 0.34% | 4.48% | −0.10% | 0.98 | 0.951 | −0.13% | 0.93 | 0.20 | −0.06 | 0.01 | 0.958 |
| | | | (−1.86) | (85.14) | | (−2.52) | (75.44) | (10.74) | (−3.16) | (0.84) | |
| 9 | 0.23% | 4.60% | −0.21% | 1.00 | 0.926 | −0.20% | 0.93 | 0.22 | −0.10 | −0.02 | 0.938 |
| | | | (−3.24) | (67.91) | | (−3.11) | (60.44) | (9.69) | (−3.80) | (−1.17) | |
| 10 (low) | 0.01% | 4.90% | −0.45% | 1.02 | 0.851 | −0.40% | 0.93 | 0.32 | −0.08 | −0.09 | 0.887 |
| | | | (−4.58) | (46.09) | | (−4.33) | (42.23) | (9.69) | (−2.23) | (−3.50) | |
| 10A | 0.25% | 4.78% | −0.19% | 1.00 | 0.864 | −0.19% | 0.91 | 0.33 | −0.11 | −0.02 | 0.891 |
| | | | (−2.05) | (48.48) | | (−2.16) | (42.99) | (10.27) | (−3.20) | (−0.76) | |
| 10B | 0.02% | 4.92% | −0.42% | 1.00 | 0.817 | −0.37% | 0.91 | 0.32 | −0.09 | −0.09 | 0.848 |
| | | | (−3.84) | (40.67) | | (−3.45) | (35.52) | (8.24) | (−2.16) | (−2.99) | |
| 10C | −0.25% | 5.44% | −0.74% | 1.05 | 0.736 | −0.64% | 0.98 | 0.32 | −0.04 | −0.17 | 0.782 |
| | | | (−5.06) | (32.16) | | (−4.49) | (28.82) | (6.29) | (−0.73) | (−4.09) | |
| 1-10 spread | 0.67% | 2.71% | 0.67% | 0.01 | −0.002 | 0.29% | −0.05 | 0.30 | 0.03 | 0.38 | 0.231 |
| | | | (4.68) | (0.39) | | (2.13) | (−1.52) | (6.30) | (0.53) | (10.07) | |
| 1A-10C spread | 1.01% | 3.87% | 1.00% | 0.02 | −0.002 | 0.53% | −0.07 | 0.40 | −0.02 | 0.50 | 0.197 |
| | | | (4.90) | (0.42) | | (2.72) | (−1.61) | (5.73) | (0.32) | (8.98) | |
| 9-10 spread | 0.22% | 1.22% | 0.23% | −0.02 | 0.004 | 0.20% | −0.01 | −0.10 | −0.01 | 0.07 | 0.118 |
| | | | (3.64) | (−1.60) | | (3.13) | (−0.40) | (−4.30) | (−0.60) | (3.87) | |

**Table 19: Carhart - Factor Loadings (Carhart, 1997, p. 64)**

| Company Name | CDS Contract | Company Name | CDS Contract |
|---|---|---|---|
| 3M | 3M COMPANY SEN 5YR CDS - CDS PREM. MID | HEWLETT-PACKARD | HEINZ (HJ) CO SEN 5YR CDS - CDS PREM. MID |
| ABBOTT LABORATORIES | ABBOTT LABORATORIES SEN 5YR CDS - CDS PREM. MID | HJ HEINZ | HEWLETT-PACKARD CO SEN 5YR CDS - CDS PREM. MID |
| ALCOA | ALCOA INC SEN 5YR CDS - CDS PREM. MID | HOME DEPOT | HOME DEPOT INC SEN 5YR CDS - CDS PREM. MID |
| ALLSTATE | ALLSTATE CORP SEN 5YR CDS - CDS PREM. MID | HONEYWELL INTL. | HONEYWELL INTL INC SEN 6YR CDS - CDS PREM. MID |
| ALTRIA GROUP | ALTRIA GROUP INC SEN 5YR CDS - CDS PREM. MID | INTEL | INTEL CORPORATION SEN 5YR CDS - CDS PREM. MID |
| AMER.ELEC.PWR. | AMERICAN ELEC PWR CO INC SEN 5YR CDS - CDS PREM. MID | INTERNATIONAL BUS.MCHS. | INTEL CORPORATION SEN 5YR CDS - CDS PREM. MID |
| AMERICAN EXPRESS | AMERICAN EXPRESS CO SEN 5YR CDS - CDS PREM. MID | JOHNSON & JOHNSON | JOHNSON & JOHNSON SEN.5YR CDS - CDS PREM. MID |
| AMGEN | AMGEN INC SEN 5YR CDS - CDS PREM. MID | JP MORGAN CHASE & CO. | JPMORGAN CHASE & CO SEN 5YR CDS - CDS PREM. MID |
| AT&T | AT&T CORP SEN 5YR CDS - CDS PREM. MID | KRAFT FOODS | KRAFT FOODS INC SEN 5YR CDS - CDS PREM. MID |
| AVON PRODUCTS | AVON PRODUCTS SEN 5YR CDS - CDS PREM. MID | LOCKHEED MARTIN | LOCKHEED MARTIN CORP SEN 5YR CDS - CDS PREM. MID |
| BAKER HUGHES | BAKER HUGHES INC SEN 5YR CDS - CDS PREM. MID | MCDONALDS | MCDONALD'S CORP SEN 5YR CDS - CDS PREM. MID |
| BANK OF AMERICA | BANK OF AMERICA CORP SEN 5YR CDS - CDS PREM. MID | MEDTRONIC | MEDTRONIC INC SEN 5YR CDS - CDS PREM. MID |
| BAXTER INTL. | BAXTER INTERNATIONAL INC SEN 5YR CDS - CDS PREM. MID | MERCK & CO. | MERCK & CO INC SEN 5YR CDS - CDS PREM. MID |
| BOEING | BOEING CAPITAL CORP SEN 5YR CDS - CDS PREM. MID | METLIFE | METLIFE INC SEN 5YR CDS - CDS PREM. MID |
| BRISTOL MYERS SQUIBB | BRISTOL-MYERS SQUIBB CO SEN 5YR CDS - CDS PREM. MID | MICROSOFT | MICROSOFT CORP SEN 5YR CDS - CDS PREM. MID |
| BURL.NTHN.SANTA FE C | BURLINGTON NORTHERN SEN 6Y CDS - CDS PREM. MID | MONSANTO | MONSANTO CO SEN 5YR CDS - CDS PREM. MID |
| CAMPBELL SOUP | CAMPBELL SOUP CO SEN 5YR CDS - CDS PREM. MID | MORGAN STANLEY | MORGAN STANLEY GP. INC SEN 5YR CDS - CDS PREM. MID |
| CAPITAL ONE FINL. | CAPITAL ONE BANK SEN 5YR CDS - CDS PREM. MID | NIKE 'B' | NIKE INCO SEN 5YR CDS - CDS PREM. MID |
| CATERPILLAR | CATERPILLAR FINANCIA SEN 6YR CDS - CDS PREM. MID | NORFOLK SOUTHERN | NORFOLK SOUTHERN CORP SEN 5YR CDS - CDS PREM. MID |
| CHEVRON | CHEVRON CORPORATION SEN 6YR CDS - CDS PREM. MID | OCCIDENTAL PTL. | OCCIDENTAL PETROLEUM CORP SEN 5YR CDS - CDS PREM. MID |
| CISCO SYSTEMS | CISCO SYSTEMS INC SEN 5YR CDS - CDS PREM. MID | ORACLE | ORACLE CORP SEN 5YR CDS - CDS PREM. MID |
| CITIGROUP | CITIGROUP INC SEN 5YR CDS - CDS PREM. MID | PEPSICO | PEPSICO INC SEN 5YR CDS - CDS PREM. MID |
| COCA COLA | COCA-COLA CO SEN 5YR CDS - CDS PREM. MID | PFIZER | PFIZER INC SEN 5YR CDS - CDS PREM. MID |
| COLGATE-PALM. | COLGATE-PALMOLIVE CO SEN 5YR CDS - CDS PREM. MID | PHILIP MORRIS INTL. | PHILIP MORRIS INTL INC SEN 5YR CDS - CDS PREM. MID |
| COMCAST 'A' | COMCAST CORP SEN 5YR CDS - CDS PREM. MID | PROCTER & GAMBLE | PROCTER & GAMBLE CO SEN.5YR CDS - CDS PREM. MID |
| CONOCOPHILLIPS | CONOCOPHILLIPS SEN 5YR CDS - CDS PREM. MID | RAYTHEON 'B' | RAYTHEON CO SEN 5YR CDS - CDS PREM. MID |
| COSTCO WHOLESALE | COSTCO WHOLESALE COR SEN 6YR CDS - CDS PREM. MID | SARA LEE | SARA LEE CORP SEN 5YR CDS - CDS PREM. MID |
| CVS CAREMARK | CVS CAREMARK CORPORA SEN 6YR CDS - CDS PREM. MID | SCHLUMBERGER | SCHLUMBERGER LTD SEN 5YR CDS - CDS PREM. MID |
| DELL | DELL INC SEN 5YR CDS - CDS PREM. MID | SOUTHERN | SOUTHERN COMPANY SEN 5YR CDS - CDS PREM. MID |
| DEVON ENERGY | DEVON ENERGY CORP SEN 5YR CDS - CDS PREM. MID | SPRINT NEXTEL | SPRINT NEXTEL CORP SEN 5YR CDS - CDS PREM. MID |
| DOW CHEMICAL | DOW CHEMICAL CO SEN 5YR CDS - CDS PREM. MID | TARGET | TARGET CORP SEN 5YR CDS - CDS PREM. MID |
| E I DU PONT DE NEMOURS | DU PONT E.I. DE NEMO URS SEN 5YR CDS - CDS PREM. MID | TEXAS INSTS. | TEXAS INSTRUMENT INC SEN 5YR CDS - CDS PREM. MID |
| ENTERGY | ENTERGY CORP SEN 5YR CDS - CDS PREM. MID | UNITED PARCEL SER. | UNITED PARCEL SER. INC SEN 5YR CDS - CDS PREM. MID |
| EXELON | EXELON CORP SEN 5YR CDS - CDS PREM. MID | UNITEDHEALTH GP. | UNITEDHEALTH GROUP INC SEN 5YR CDS - CDS PREM. MID |
| EXXON MOBIL | EXXON MOBIL CORP SEN 5YR CDS - CDS PREM. MID | US BANCORP | U.S. BANCORP SEN 5YR CDS - CDS PREM. MID |
| FEDEX | FEDEX CORP SEN 5YR CDS - CDS PREM. MID | VERIZON COMMUNICATIONS | VERIZON WIRELESS INC SEN 5YR CDS - CDS PREM. MID |
| FORD MOTOR | FORD MOTOR CO SEN 5YR CDS - CDS PREM. MID | WAL MART STORES | WAL-MART STORES INC SEN 5YR CDS - CDS PREM. MID |
| FREEPORT-MCMOR.CPR.& GD. | FREEPORT-MCMOR.CPR.& GD. SEN 5YR CDS - CDS PREM. MID | WALT DISNEY | WALT DISNEY CO/THE SEN 5YR CDS - CDS PREM. MID |
| GENERAL DYNAMICS | GENERAL DYNAMICS CORP SEN 5YR CDS - CDS PREM. MID | WELLS FARGO & CO | WELLS FARGO & CO SEN 5YR CDS - CDS PREM. MID |
| GENERAL ELECTRIC | GENERAL ELEC.CAPITAL CORP SEN 5YR CDS - CDS PREM. MID | WEYERHAEUSER | WEYERHAEUSER CO SEN 5YR CDS - CDS PREM. MID |
| GOLDMAN SACHS GP. | GOLDMAN SACHS GP INC SEN 5YR CDS - CDS PREM. MID | XEROX | XEROX CORP SEN 5YR CDS - CDS PREM. MID |
| HALLIBURTON | HALLIBURTON CO SEN 5YR CDS - CDS PREM. MID | | |
| **N = 42** | | **N = 41** | |

**Table 20: Stock Universe**

MB Deciles regressed against RMU and RM

| | MB Deci~1 | 2 | 3 | 4 | 5 | 6 | 7 | 8 | 9 | MB Dec~10 |
|---|---|---|---|---|---|---|---|---|---|---|
| rmu | 0.964*** | 0.370*** | 0.000151 | -0.0278 | -0.149*** | -0.184*** | -0.231*** | -0.226*** | -0.256*** | -0.295*** |
| | (24.44) | (10.20) | (0.00) | (-1.16) | (-5.67) | (-8.04) | (-8.51) | (-9.37) | (-12.00) | (-15.11) |
| rm | 1.011*** | 1.006*** | 1.148*** | 1.021*** | 0.993*** | 0.990*** | 1.010*** | 0.986*** | 0.934*** | 0.900*** |
| | (28.81) | (31.55) | (36.17) | (47.67) | (38.98) | (40.20) | (44.09) | (40.66) | (38.88) | (51.62) |
| _cons | -0.000572** | -0.000108 | -0.0000704 | 0.0000394 | 0.0000406 | 0.0000322 | 0.000138 | 0.000247 | 0.000267 | 0.000184 |
| | (-2.67) | (-0.57) | (-0.41) | (0.27) | (0.28) | (0.22) | (1.01) | (1.76) | (1.91) | (1.48) |
| N | 1304 | 1304 | 1304 | 1304 | 1304 | 1304 | 1304 | 1304 | 1304 | 1304 |
| R-sq | 0.926 | 0.891 | 0.890 | 0.893 | 0.870 | 0.860 | 0.874 | 0.864 | 0.843 | 0.855 |

t statistics in parentheses
* p<0.05, ** p<0.01, *** p<0.001

**Table 21: MB deciles against RMU and RM**

MB Deciles regressed against VMC and RM

| | MB Deci~1 | 2 | 3 | 4 | 5 | 6 | 7 | 8 | 9 | MB Dec~10 |
|---|---|---|---|---|---|---|---|---|---|---|
| vmc | 0.443*** | 0.0996*** | 0.000520 | -0.0139 | -0.0682*** | -0.0895*** | -0.106*** | -0.0847*** | -0.0970*** | -0.106*** |
| | (23.10) | (5.20) | (0.03) | (-1.07) | (-5.31) | (-8.80) | (-8.30) | (-6.47) | (-7.67) | (-9.87) |
| rm | 1.112*** | 1.147*** | 1.158*** | 1.019*** | 0.978*** | 0.970*** | 1.003*** | 0.944*** | 0.877*** | 0.823*** |
| | (23.13) | (23.56) | (29.39) | (37.01) | (32.64) | (36.63) | (36.50) | (28.53) | (27.16) | (32.44) |
| _cons | 0.00122* | 0.000405 | -0.000444 | -0.0000137 | -0.000174 | -0.000400 | -0.000188 | -0.0000793 | -0.0000529 | -0.000272 |
| | (2.47) | (0.75) | (-1.03) | (-0.04) | (-0.53) | (-1.19) | (-0.62) | (-0.26) | (-0.15) | (-0.95) |
| N | 435 | 435 | 435 | 435 | 435 | 435 | 435 | 435 | 435 | 435 |
| R-sq | 0.951 | 0.890 | 0.911 | 0.920 | 0.901 | 0.891 | 0.910 | 0.899 | 0.858 | 0.879 |

t statistics in parentheses
* p<0.05, ** p<0.01, *** p<0.001

**Table 22: MB deciles against VMC and RM**

MB Deciles regressed against RMU, VMC, and RM

| | MB Deci~1 | 2 | 3 | 4 | 5 | 6 | 7 | 8 | 9 | MB Dec~10 |
|---|---|---|---|---|---|---|---|---|---|---|
| rmu | 0.523*** | 0.465*** | -0.0710 | -0.000227 | -0.0906* | -0.0740 | -0.179*** | -0.132** | -0.195*** | -0.257*** |
| | (7.97) | (6.34) | (-1.15) | (-0.01) | (-2.04) | (-1.48) | (-3.97) | (-2.67) | (-4.84) | (-7.21) |
| vmc | 0.265*** | -0.0585 | 0.0247 | -0.0138 | -0.0373* | -0.0643** | -0.0450** | -0.0397 | -0.0307 | -0.0190 |
| | (8.33) | (-1.75) | (0.91) | (-0.68) | (-2.08) | (-3.18) | (-3.04) | (-1.87) | (-1.50) | (-1.16) |
| rm | 0.997*** | 1.045*** | 1.173*** | 1.019*** | 0.998*** | 0.987*** | 1.042*** | 0.973*** | 0.920*** | 0.880*** |
| | (25.81) | (24.24) | (27.50) | (35.18) | (30.83) | (33.19) | (35.92) | (29.48) | (27.93) | (38.20) |
| _cons | 0.000393 | -0.000331 | -0.000332 | -0.0000133 | -0.0000307 | -0.000283 | 0.0000954 | 0.000130 | 0.000255 | 0.000135 |
| | (0.89) | (-0.67) | (-0.77) | (-0.04) | (-0.09) | (-0.89) | (0.33) | (0.43) | (0.72) | (0.49) |
| N | 435 | 435 | 435 | 435 | 435 | 435 | 435 | 435 | 435 | 435 |
| R-sq | 0.963 | 0.908 | 0.912 | 0.920 | 0.903 | 0.892 | 0.917 | 0.903 | 0.868 | 0.901 |

t statistics in parentheses
* p<0.05, ** p<0.01, *** p<0.001

**Table 23: MB deciles against RMU, VMC, and RM**

MB Quintiles regressed against RMU, VMC, SMB, HML and RM

| | MB Quin~1 | 2 | 3 | 4 | MB Quin~5 |
|---|---|---|---|---|---|
| rmu | 0.378*** | -0.0416 | -0.101* | -0.114*** | -0.143** |
| | (6.19) | (-0.86) | (-2.38) | (-3.42) | (-3.30) |
| vmc | 0.113*** | 0.00886 | -0.0475** | -0.0437*** | -0.0298* |
| | (4.23) | (0.44) | (-3.24) | (-3.60) | (-2.06) |
| smb | 0.0122 | -0.0808 | 0.0765* | 0.0395 | -0.0495 |
| | (0.25) | (-1.63) | (2.06) | (1.27) | (-1.45) |
| hml | 0.284*** | -0.0119 | 0.0204 | -0.105** | -0.182** |
| | (4.97) | (-0.23) | (0.33) | (-2.78) | (-2.67) |
| rm | 1.033*** | 1.095*** | 0.992*** | 1.011*** | 0.889*** |
| | (41.90) | (38.67) | (41.08) | (57.38) | (45.25) |
| _cons | 0.0000940 | -0.000149 | -0.000161 | 0.0000266 | 0.000213 |
| | (0.26) | (-0.48) | (-0.64) | (0.13) | (0.87) |
| N | 367 | 367 | 367 | 367 | 367 |
| R-sq | 0.975 | 0.959 | 0.951 | 0.964 | 0.947 |

t statistics in parentheses
* p<0.05, ** p<0.01, *** p<0.001

**Table 24: MB deciles against RMU, VMC, SMB, HML, and RM**

. dfuller rmu

Dickey-Fuller test for unit root Number of obs = 434

| | Test Statistic | 1% Critical Value | 5% Critical Value | 10% Critical Value |
|---|---|---|---|---|
| | | Interpolated Dickey-Fuller | | |
| Z(t) | -20.282 | -3.445 | -2.873 | -2.570 |

MacKinnon approximate p-value for Z(t) = 0.0000

. dfuller hml

Dickey-Fuller test for unit root Number of obs = 354

| | Test Statistic | 1% Critical Value | 5% Critical Value | 10% Critical Value |
|---|---|---|---|---|
| | | Interpolated Dickey-Fuller | | |
| Z(t) | -19.427 | -3.452 | -2.876 | -2.570 |

MacKinnon approximate p-value for Z(t) = 0.0000

**Table 25: Dickey-Fuller Test on RMU and HML**